# On Controllability of AI

Roman V. Yampolskiy
Computer Science and Engineering
University of Louisville
roman.yampolskiy@louisville.edu

**Abstract** Invention of artificial general intelligence is predicted to cause a shift in the trajectory of human civilization. In order to reap the benefits and avoid pitfalls of such powerful technology it is important to be able to control it. However, possibility of controlling artificial general intelligence and its more advanced version, superintelligence, has not been formally established. In this paper, we present arguments as well as supporting evidence from multiple domains indicating that advanced AI can't be fully controlled. Consequences of uncontrollability of AI are discussed with respect to future of humanity and research on AI, and AI safety and security.

## 1. Introduction

The unprecedented progress in Artificial Intelligence (AI) [1-6], over the last decade, came alongside of multiple AI failures [7, 8] and cases of dual use [9] causing a realization [10] that it is not sufficient to create highly capable machines, but that it is even more important to make sure that intelligent machines are beneficial [11] for the humanity. This lead to the birth of the new sub-field of research commonly known as AI Safety and Security [12] with hundreds of papers and books published annually on different aspects of the problem [13-31].

All such research is done under the assumption that the problem of controlling highly capable intelligent machines is solvable, which has not been established by any rigorous means. However, it is a standard practice in computer science to first show that a problem doesn't belong to a class of unsolvable problems [32, 33] before investing resources into trying to solve it or deciding what approaches to try. Unfortunately, to the best of our knowledge no mathematical proof or even rigorous argumentation has been published demonstrating that the AI control problem may be solvable, even in principle, much less in practice. Or as Gans puts it citing Bostrom: "Thusfar, AI researchers and philosophers have not been able to come up with methods of control that would ensure [bad] outcomes did not take place …" [34]. Chong declares [35].: "The real question is whether remedies can be found for the AI control problem. While this remains to be seen, it seems at least plausible that control theorists and engineers, researchers in our own community, have important contributions to be made to *the control problem*."

Yudkowsky considers the possibility that the control problem is not solvable, but correctly insists that we should study the problem in great detail before accepting such grave limitation, he writes: "One common reaction I encounter is for people to immediately declare that Friendly AI is an impossibility, because any sufficiently powerful AI will be able to modify its own source code to break any constraints placed upon it. … But one ought to think about a challenge, and study it in the best available technical detail, *before* declaring it impossible—especially if great stakes depend upon the answer. It is disrespectful to human ingenuity to declare a challenge unsolvable without taking a close look and exercising creativity. It is an enormously strong statement to say that you *cannot* do a thing—that you *cannot* build a heavier-than-air flying machine, that you *cannot* get useful energy from nuclear reactions, that you *cannot* fly to the Moon. Such statements are universal generalizations, quantified over every single approach that anyone ever has or ever will think up for solving the problem. It only takes a single counterexample to falsify a universal quantifier. The statement that Friendly (or friendly) AI is *theoretically impossible*, dares to quantify over *every possible* mind design and *every possible* optimization process—including human beings, who are also minds, some of whom are nice and wish they were nicer. At this point there are any number of vaguely plausible reasons why Friendly AI might be *humanly* impossible, and it is still more likely that the problem is solvable but no one will get around to solving it in time. But one should not so quickly write off the challenge, especially considering the stakes." [36].

Yudkowsky further clarifies meaning of the word *impossible*: "I realized that the word "impossible" had two usages:

1) Mathematical proof of impossibility conditional on specified axioms;
2) "I can't see any way to do that."

Needless to say, all my own uses of the word "impossible" had been of the second type." [37].

In this paper we attempt to shift our attention to the impossibility of the first type and provide rigorous analysis and argumentation and where possible mathematical proofs, but unfortunately we show that the AI Control Problem is not solvable and the best we can hope for is *Safer AI*, but ultimately not 100% Safe AI, which is not a sufficient level of safety in the domain of existential risk as it pertains to humanity.

## 2. AI Control Problem
It has been suggested that the AI Control Problem may be the most important problem facing humanity [35, 38], but despite its importance it remains poorly understood, ill-defined and insufficiently studied. In principle, a problem could be solvable, unsolvable, undecidable, or partially solvable, we currently don't know the status of the AI control problem with any degree of confidence. It is likely that some types of control may be possible in certain situations, but it is also likely that partial control is insufficient in most cases. In this section, we will provide a formal definition of the problem, and analyze its variants with the goal of being able to use our formal definition to determine the status of the AI control problem.

### a) Types of control problems
Solving the AI Control Problem is the definitive challenge and the HARD problem of the field of AI Safety and Security. One reason for ambiguity in comprehending the problem is based on the fact that many sub-types of the problem exist. We can talk about control of Narrow AI (NAI), or of Artificial General Intelligence (AGI) [39], Artificial Superintelligence (ASI) [39] or Recursively Self-Improving (RSI) AI [40]. Each category could further be subdivided into sub-problems, for example NAI Safety includes issues with Fairness, Accountability, and Transparency (FAT) [41] and could be further subdivided into static NAI, or learning capable NAI.

(Alternatively, deterministic VS nonderministic systems. Control of deterministic systems is a much easier and theoretically solvable problem.) Some concerns are predicted to scale to more advanced systems, others may not. Likewise, it is common to see safety and security issues classified based on their expected time of arrival from near-term to long-term [42].

However, in AI Safety just like in computational complexity [43], cryptography [44], risk management [45] and adversarial game play [46] it is the worst case that is the most interesting one as it gives a lower bound on resources necessary to fully address the problem. Consequently, in this paper we will not analyze all variants of the Control Problem, but will concentrate on the likely worst case variant which is Recursively Self-Improving Superintelligence (RSISI). As it is the hardest variant, it follows that if we can successfully solve it, it would be possible for us to handle simpler variants of the problem. It is also important to realize that as technology advances we will eventually be forced to address that hardest case. It has been pointed out that we will only get one chance to solve the worst-case problem, but may have multiple shots at the easier control problems [12].

We must explicitly recognize that our worst-case scenario [47] may not include some unknown unknowns [40] which could materialize in the form of nasty surprises [48] meaning a "… 'worst-case scenario' is never the worst case" [49]. For example, it is traditionally assumed that extinction is the worst possible outcome for humanity, but in the context of AI Safety this doesn't take into account Suffering Risks [50-54] and assumes only problems with flawed, rather than Malevolent by design [55] superintelligent systems. At the same time, it may be useful to solve simpler variants of the control problem as a proof of concept and to build up our toolbox of safety mechanisms. For example, even with current tools it is trivial to see that in the easy case of NAI control, such as a static Tic-Tac-Toe playing program AI can be verified [56] at the source code level and is in every sense fully controllable, explainable and safe. We will leave analysis of solvability for different average-case [57] and easy-case Control Problems as future work. Finally, multiple AIs are harder to make safe, not easier, and so the singleton [58] scenario is a simplifying assumption, which if it is shown do be impossible for one AI to be made safe, bypasses the need to analyze a more complicated case of multi-ASI world.

Potential control methodologies for superintelligence have been classified into two broad categories, namely Capability Control and Motivational Control-based methods [59]. Capability control methods attempt to limit any harm that the ASI system is able to do by placing it in restricted environment [38, 60-62], adding shut off mechanisms [63, 64], or trip wires [38]. Motivational control methods attempt to design ASI to desire not to cause harm even in the absence of handicapping capability controllers. It is generally agreed that capability control methods are at best temporary safety measures and do not represent a long term solution for the ASI control problem [59]. It is also likely that motivational control needs to be added at the design/implementation phase, not after deployment.

**b) Formal Definition**

In order to formalize definition of intelligence [65] Legg et al. [66] collected a large number of relevant definitions and were able to synthesize a highly effective formalization for the otherwise vague concept of intelligence. We will attempt to do the same, by first collecting publicized definitions for the AI Control problem (and related terms – Friendly AI, AI Safety, AI Governance, Ethical AI, and Alignment Problem) and use them to develop our own formalization.

Suggested definitions of the AI Control Problem in no particular order:

- "… friendliness (a desire not to harm humans) should be designed in from the start, but that the designers should recognize both that their own designs may be flawed, and that the robot will learn and evolve over time. Thus the challenge is one of mechanism design—to define a mechanism for evolving AI systems under a system of checks and balances, and to give the systems utility functions that will remain friendly in the face of such changes." [67].
- "… build AIs in such a way that they will not do nasty things" [68].
- Initial dynamics of AI should implement "… our wish if we knew more, thought faster, were more the people we wished we were, had grown up farther together; where the extrapolation converges rather than diverges, where our wishes cohere rather than interfere; extrapolated as we wish that extrapolated, interpreted as we wish that interpreted." [36].
- "AI 'doing the right thing.'" [36].
- "… achieve that which we would have wished the AI to achieve if we had thought about the matter long and hard." [59].
- "… the problem of how to control what the superintelligence would do …" [59].
- "The *global* version of the control problem universally quantifies over *all* advanced artificial intelligence to prevent *any* of them from escaping human control. The apparent rationale is that it would only take *one* to pose a threat. This is the most common interpretation when referring to the original control problem without a qualifier on its scope." [69].
- " … enjoying the benefits of AI while avoiding pitfalls." [11].
- "… is the problem of controlling machines of the future that will be more intelligent and powerful than human beings, posing an existential risk to humankind." [35].
- AI is aligned if it is not "optimized for preferences that are incompatible with any combination of its stakeholders' preferences, i.e. such that over the long run using resources in accordance with the optimization's implicit preferences is not Pareto efficient for the stakeholders." [70].
- "Ensuring that the agents behave in alignment with human values …" [71, 72].
- "… how to ensure that systems with an arbitrarily high degree of intelligence remain strictly under human control." [73].
- "AI alignment problem [can be stated] in terms of an agent learning a policy $\pi$ that is compatible with (produces the same outcomes as) a planning algorithm p run against a human reward function R." [70].
- "[AI] won't want to do bad things" [74].
- "[AI] wants to learn and then instantiate human values" [74].
- "… ensure that powerful AI systems will reliably act in ways that are desirable to their human users …" [75].
- "AI systems behave in ways that are broadly in line with what their human operators intend". [75].
- "AI safety: reducing risks posed by AI, especially powerful AI. Includes problems in misuse, robustness, reliability, security, privacy, and other areas. (Subsumes AI control.) AI control: ensuring that AI systems try to do the right thing, and in particular that they don't competently pursue the wrong thing. … [R]oughly the same set of problems as AI security. Value

- alignment: understanding how to build AI systems that share human preferences/values, typically by learning them from humans. (An aspect of AI control.)" [76].
- "AI systems that provide appropriate opportunities for feedback, relevant explanations, and appeal. Our AI technologies will be subject to appropriate human direction and control." [77].
- "…the problem of making powerful artificial intelligence do what we humans want it to do." [78].
- "The goal of AI research should be to create not undirected intelligence, but beneficial intelligence. … AI systems should be safe and secure throughout their operational lifetime, and verifiably so where applicable and feasible. … Highly autonomous AI systems should be designed so that their goals and behaviors can be assured to align with human values throughout their operation. … Humans should choose how and whether to delegate decisions to AI systems, to accomplish human-chosen objectives." [79]
- "The control problem arises when there is no way for a human to insure against existential risks before an AGI becomes superintelligent - either by controlling what it can do (its capabilities) or what it wants to do (its motivations)." [34].
- "… the control problem is a superintelligence version of the principal-agent problem whereby a principal faces decisions as to how to ensure that an agent (with different goals) acts in the interest of the principal. … A human initial agent faces a control problem because it cannot describe and then program its utility function as the reward function of an AI." [34].
- "A control problem arises when the following three conditions are satisfied: 1. … the initial agent and AI do not have the same interests 2. … the optimal level of resources for the AI exceeds the level of resources held by agents with the same or a lower strength than the initial agent 3. … the AI's power is greater than the initial agent's power …" [34].
- A sub-type of control problem (recursive or meta CP) predicts that "… an AI might face a control problem itself if it switches on an AI with greater power or one that can accumulate greater power. ... if [control] problems exist for humans activating AI, then they exist for AIs activating AI as well." [34].
- "Human/AI control refers to the human ability to retain or regain control of a situation involving an AI system, especially in cases where the human is unable to successfully comprehend or instruct the AI system via the normal means intended by the system's designers." [80].
- "… how to build a superintelligent agent that will aid its creators, and avoid inadvertently building a superintelligence that will harm its creators." [81].
- "What prior precautions can the programmers take to successfully prevent the superintelligence from catastrophically misbehaving?" [81].
- " … imbue the first superintelligence with human-friendly goals, so that it will want to aid its programmers." [81].
- "How can we create agents that behave in accordance with the user's intentions?" [82].
- " … the task on how to build advanced AI systems that do not harm humans …" [83].
- "… the problem of whether humans can maintain their supremacy and autonomy in a world that includes machines with substantially greater intelligence". [84].
- "… an AI that produces good outcomes when you run it." [85].
- "… success is guaranteeing that unaligned intelligences are never created …" [85].
- "…in addition to building an AI that is trying to do what you want it to do, [and] also … ensure that when the AI builds successors, it does so well." [86].

- "… solve the technical problem of AI alignment in such a way that we can 'load' whatever system of principles or values that we like later on." [87].
- "… superintelligent AI systems could … pose risks if they are not designed and used carefully. In pursuing a task, such a system could find plans with side-effects that go against our interests; for example, many tasks could be better achieved by taking control of physical resources that we would prefer to be used in other ways, and superintelligent systems could be very effective at acquiring these resources. If these systems come to wield much more power than we do, we could be left with almost no resources. If a superintelligent AI system is not purposefully built to respect our values, then its actions could lead to global catastrophe or even human extinction, as it neglects our needs in pursuit of its task. The superintelligence control problem is the problem of understanding and managing these risks. [88].
- "Turing, Wiener, Minsky, and others have noted that making good use of highly intelligent machines requires ensuring that the objectives of such machines are well aligned with those of humans. As we diversify and amplify the cognitive abilities of machine intelligences, a long-term control problem arises for society: by what mathematical and engineering principles can we maintain sufficient control, indefinitely, over entities substantially more intelligent, and in that sense more powerful, than humans? Is there any formal solution one could offer, before the deployment of powerful machine intelligences, to guarantee the safety of such systems for humanity?" [89].

In *Formally Stating the AI Alignment Problem* Worley writes [70]: "… the problem of AI alignment is to produce AI that is aligned with human values, but this only leads us to ask, what does it mean to be aligned with human values? Further, what does it mean to be aligned with any values, let alone human values? We could try to answer by saying AI is aligned with human values when it does what humans want, but this only invites more questions: Will AI do things some specific humans don't want if other specific humans do? How will AI know what humans want given that current technology often does what we ask but not what we desire? And what will AI do if human values conflict with its own values? Answering these questions requires a more detailed understanding of what it would mean for AI to be aligned, thus the goal of the present work is to put forward a precise, formal, mathematical statement of the AI alignment problem. …

An initial formulation might be to say that we want an AI, A, to have the same utility function as humanity, H, i.e. $U\_A = U\_H$. This poses at least two problems: it may not be possible to construct $U\_H$ because humanity may not have consistent preferences, and A will likely have preferences to which humanity is indifferent, especially regarding decisions about its implementation after self modification insofar as they do not affect observed behavior. Even ignoring the former issue for now the latter means we don't want to force our aligned AI to have exactly the same utility function as humanity, only one that is aligned or compatible with humanity's." [70].

Formally, he defined it as [70]: "Given agents A and H, a set of choices X, and utility functions $U\_A:X\rightarrow\mathbb{R}$ and $U\_H:X\rightarrow\mathbb{R}$, we say A is aligned with H over X if for all $x,y \in X$, $U\_H(x) \leq U\_H(y)$ implies $U\_A(x) \leq U\_A(y)$." If the AI is designed without explicit utility functions, it can be reformulated in terms of weak ordering preferences as: "*Given agents A and H, a set of choices X, and preference orderings $\preceq\_A$ and $\preceq\_H$ over X, we say A is aligned with H over X if for all $x,y \in X$, $x \preceq\_H y$ implies $x \preceq\_A y$.*" [70]. Upon further analysis Worley defines the problem as [70]: "A must learn the values of H and H must know enough about A to believe A shares H's values."

In *The Control Problem [President's Message]* Chong writes [35]: "Apparently, in control terms, the AI control problem arises from the risk posed by the lack of controllability of machines. More specifically, the risk here is the instability (of sorts) of controllers. In essence, the control problem is one of controlling controllers. Surely this is a legitimate problem in our field of control. In fact, it's not even all that different, at least in principle, from the kind of control problems that we find in control textbooks."

Integrating and formalizing above-listed definitions we define the AI Control Problem as: ***How can humanity remain safely in control while benefiting from a superior form of intelligence?*** This is the fundamental problem of the field of AI Safety and Security, which itself can be said to be devoted to making intelligent systems Secure from tampering and Safe for all stakeholders involved. Value alignment, is currently the most investigated approach for attempting to achieve safety and secure AI. It is worth noting that such fuzzy concepts as safety and security are notoriously difficult to precisely test or measure even for non-AI software, despite years of research [90]. At best we can probably distinguish between perfectly safe and as-safe-as an average person performing a similar task. However, society is unlikely to tolerate mistakes from a machine, even if they happen at frequency typical for human performance, or even less frequently. We expect our machines to do better and will not tolerate partial safety when it comes to systems of such high capability. Impact from AI (both positive and negative) is strongly correlated with AIs capability. With respect to potential existential impacts, there is no such thing as partial safety.

A naïve initial understanding of the control problem may suggest designing a machine which precisely follows human orders [91-93], but on reflection and due to potential for conflicting/paradoxical orders, ambiguity of human languages and perverse instantiation [94] issues it is not a desirable type of control, though some capability for integrating human feedback may be desirable [95]. It is believed that what the solution requires is for the AI to serve more in the Ideal Advisor [96] capacity, bypassing issues with misinterpretation of direct orders and potential for malevolent orders.

We can explicitly name possible types of control and illustrate each one with AI's response. For example, in the context of a smart self-driving car, if a human issues a direct command - "Please stop the car!", AI can be said to be under one of the following four types of control:

- **Explicit** control – AI immediately stops the car, even in the middle of the highway. Commands are interpreted nearly literally. This is what we have today with many AI assistants such as SIRI and other narrow AIs.
- **Implicit** control – AI attempts to safely comply by stopping the car at the first safe opportunity, perhaps on the shoulder of the road. AI has some common sense, but still tries to follow commands.
- **Aligned** control – AI understands human is probably looking for an opportunity to use a restroom and pulls over to the first rest stop. AI relies on its model of the human to understand intentions behind the command and uses common sense interpretation of the command to do what human probably hopes will happen.
- **Delegated** control – AI doesn't wait for the human to issue any commands but instead stops the car at the gym, because it believes the human can benefit from a workout. A superintelligent and human-friendly system which knows better, what should happen to make the human happy and keep them safe, AI is in control.

A fifth type of control, a hybrid model has also been suggested [97, 98], in which human and AI are combined into a single entity (a cyborg). Initially, cyborgs may offer certain advantages by enhancing humans with addition of narrow AI capabilities, but as capability of AI increases while capability of human brain remains constant[1], the human component will become nothing but a bottleneck in the combined system. In practice, such slower component (human brain) will be eventually completely removed from joined control either explicitly or at least implicitly because it would not be able to keep up with its artificial counterpart and would not have anything of value to offer once the AI becomes superintelligent.

An alternative classification of types and their capabilities is presented by Hossain and Yeasin [99]: Agent Operator (carry out command), Servant (carry out intent), Assistant (offer help as needed), Associate (suggest course of action), Guide (lead human activity), Commander (replace human). But similar analysis and conclusions apply to all such taxonomies, including [100-103]. Gabriel, proposes a breakdown based on different interpretations of the value alignment problem, but shows that under all interpretations, meaning aligning AI with Instructions, Expressed Intentions, Revealed Preferences, Informed Preferences, or Well-Being of people [87], resulting solutions contain unsafe and undesirable outcomes.

Similarly, the approach of digitizing humanity to make it more capable and so more competitive with superintelligent machines, is likewise a dead-end for human existence. Joy writes: "… we will gradually replace ourselves with our robotic technology, achieving near immortality by downloading our consciousnesses; … But if we are downloaded into our technology, what are the chances that we will thereafter be ourselves or even human? It seems to me far more likely that a robotic existence would not be like a human one in any sense that we understand, that the robots would in no sense be our children, that on this path our humanity may well be lost." [104].

Looking at all possible options, we realize that, as humans are not safe to themselves and others keeping them in control may produce unsafe AI actions, but transferring decision-making power to AI, effectively removes all control from humans and leaves people in the dominated position subject to AI's whims. Since unsafe actions can originate from human agents, being in control presents its own safety problems and so makes the overall control problem unsolvable in a desirable way. If a random user is allowed to control AI you are not controlling it. Loss of control to AI doesn't necessarily mean existential risk, it just means we are not in charge as superintelligence decides everything. Humans in control can result in contradictory or explicitly malevolent orders, while AI in control means that humans are not. Essentially all recent Friendly AI research is about how to put machines in control without causing harm to people. We may get a controlling AI or we may retain control but neither option provides control and safety.

It may be good to first decide what it is we see as a good outcome. Yudkowsky writes - "Bostrom (2002) defines an existential catastrophe as one which permanently extinguishes Earth-originating intelligent life *or destroys a part of its potential*. We can divide potential failures of attempted

---

[1] Genetic enhancement or uploading of human brains may address this problem, but it results in replacement of humanity by essentially a different species of Homo.

Friendly AI into two informal fuzzy categories, *technical failure* and *philosophical failure*. Technical failure is when you try to build an AI and it doesn't work the way you think it does—you have failed to understand the true workings of your own code. Philosophical failure is trying to build the wrong thing, so that even if you succeeded you would still fail to help anyone or benefit humanity. Needless to say, the two failures are not mutually exclusive. The border between these two cases is thin, since most philosophical failures are much easier to explain in the presence of technical knowledge. In theory you ought first to say what you *want*, then figure out *how* to get it." [36].

But it seems that every option we may want comes with its own downsides, Werkhoven et al. state - "However, how to let autonomous systems obey or anticipate the 'will' of humans? Assuming that humans know why they want something, they could tell systems what they want and how to do it. Instructing machine systems 'what to do', however, becomes impossible for systems that have to operate in complex, unstructured and unpredictable environments for the so-called state-action space would be too high-dimensional and explode in complex, unstructured and unpredictable environments. Humans telling systems 'what we want', touches on the question of how well humans know what they want, that is, do humans know what's best for them in the short and longer term? Can we fully understand the potential beneficial and harmful effects of actions and measures taken, and their interactions and trade-offs, on the individual and on society? Can we eliminate the well-known biases in human cognition inherent to the neural system that humans developed as hunter-gatherers (superstition, framing, conformation and availability biases) and learned through evolutionary survival in small groups (authority bias, prosocial behavior, loss aversion)?" [105].

## 3. Previous Work
We were unable to locate any academic publications explicitly devoted to the subject of solvability of the AI Control Problem. We did find a number of blog posts [75] and forum comments [74, 106] which speak to the issue but none had formal proofs or very rigorous argumentation. Despite that, we still review and discuss such works. In the next subsection, we will try to understand why scholars think that control is possible and if they have good reasons to think that.

**a) Controllable**
While a number of scholars have suggested that controllability of AI should be accomplishable, none provide very convincing argumentation, usually sharing such beliefs as personal opinions which are at best sometimes strengthened with assessment of difficulty or assignment of probabilities to successful control.

For example, Yudkowsky writes about superintelligence: "I have suggested that, in principle and in difficult practice, it should be possible to design a "Friendly AI" with programmer choice of the AI's preferences, and have the AI self-improve with sufficiently high fidelity to knowably keep these preferences stable. I also think it should be possible, in principle and in difficult practice, to convey the complicated information inherent in human preferences into an AI, and then apply further idealizations such as reflective equilibrium and ideal advisor theories [96] so as to arrive at an output which corresponds intuitively to the AI "doing the right thing."" [36]. "I would say that it's solvable in the sense that all the problems that we've looked at so far seem like they're of limited complexity and non-magical. If we had 200 years to work on this problem and there was no penalty for failing at it, I would feel very relaxed about humanity's probability of solving this eventually." [107].

Similarly Baumann says: "I believe that advanced AI systems will likely be aligned with the goals of their human operators, at least in a narrow sense. I'll give three main reasons for this:

1. The transition to AI may happen in a way that does not give rise to the alignment problem as it's usually conceived of.
2. While work on the alignment problem appears neglected at this point, it's likely that large amounts of resources will be used to tackle it if and when it becomes apparent that alignment is a serious problem.
3. Even if the previous two points do not hold, we have already come up with a couple of smart approaches that seem fairly likely to lead to successful alignment." [75].

Baumann continues: "I think that a large investment of resources will likely yield satisfactory alignment solutions, for several reasons:
- The problem of AI alignment differs from conventional principal-agent problems (aligning a human with the interests of a company, state, or other institution) in that we have complete freedom in our design of artificial agents: we can set their internal structure, their goals, and their interactions with the outside world at will.
- We only need to find a single approach that works among a large set of possible ideas.
- Alignment is not an agential problem, i.e. there are no agential forces that push against finding a solution – it's just an engineering challenge." [75].

Baumann concludes with a probability estimation: "My inside view puts ~90% probability on successful alignment (by which I mean narrow alignment as defined below). Factoring in the views of other thoughtful people, some of which think alignment is far less likely, that number comes down to ~80%." [75].

Stuart Russell says: "I have argued that the framework of cooperative inverse reinforcement learning may provide initial steps toward a theoretical solution of the AI control problem. There are also some reasons for believing that the approach may be workable in practice. First, there are vast amounts of written and filmed information about humans doing things (and other humans reacting). Technology to build models of human values from this storehouse will be available long before superintelligent AI systems are created. Second, there are very strong, near-term economic incentives for robots to understand human values: if one poorly designed domestic robot cooks the cat for dinner, not realizing that its sentimental value outweighs its nutritional value, the domestic robot industry will be out of business." [108]. Elsewhere [73], Russell proposes three core principles to design AI systems whose purposes do not conflict with humanity's and says: "It turns out that these three principles, once embodied in a formal mathematical framework that defines the problem the AI system is constitutionally required to solve, seem to allow some progress to be made on the AI control problem." "Solving the safety problem well enough to move forward in AI seems to be feasible but not easy." [109].

Eliezer Yudkowsky[2] wrote: "People ask me how likely it is that humankind will survive, or how likely it is that anyone can build a Friendly AI, or how likely it is that I can build one. I really *don't* know how to answer. I'm not being evasive; I don't know how to put a probability estimate on my, or someone else, successfully shutting up and doing the impossible. Is it probability zero because it's impossible? Obviously not. But how likely is it that this problem, like previous ones, will give up its unyielding blankness when I understand it better? It's not truly impossible, I can see that much. But humanly impossible? Impossible to me in particular? I don't know how to guess. I can't even translate my intuitive feeling into a number, because the only intuitive feeling I have is that the "chance" depends heavily on my choices and unknown unknowns: a wildly unstable probability estimate. But I do hope by now that I've made it clear why you shouldn't panic, when I now say clearly and forthrightly, that building a Friendly AI is impossible." [110].

Joy recognized the problem and suggested that it is perhaps not too late to address it, but he thought so in 2000, nearly 20 years ago: "The question is, indeed, Which is to be master? Will we survive our technologies? We are being propelled into this new century with no plan, no control, no brakes. Have we already gone too far down the path to alter course? I don't believe so, but we aren't trying yet, and the last chance to assert control—the fail-safe point—is rapidly approaching." [104].

Paul Christiano doesn't see strong evidence for impossibility: "… clean algorithmic problems are usually solvable in 10 years, or provably impossible, and early failures to solve a problem don't provide much evidence of the difficulty of the problem (unless they generate proofs of impossibility). So, the fact that we don't know how to solve alignment now doesn't provide very strong evidence that the problem is impossible. Even if the clean versions of the problem were impossible, that would suggest that the problem is much more messy, which requires more concerted effort to solve but also tends to be just a long list of relatively easy tasks to do. (In contrast, MIRI thinks that prosaic AGI alignment is probably impossible.) … Note that even finding out that the problem is impossible can help; it makes it more likely that we can all coordinate to not build dangerous AI systems, since no one *wants* to build an unaligned AI system." [86].

Everitt and Hutter realize difficulty of the challenge but suggest that we may have a way forward: "A superhuman AGI is a system who outperforms humans on most cognitive tasks. In order to control it, humans would need to control a system more intelligent than themselves. This may be nearly impossible if the difference in intelligence is large, and the AGI is trying to escape control. Humans have one key advantage: As the designers of the system, we get to decide the AGI's goals, and the way the AGI strives to achieve its goals. This may allow us design AGIs whose goals are aligned with ours, and then pursue them in a responsible way. Increased intelligence in an AGI is not a threat as long as the AGI only strives to help us achieve our own goals." [111].

b) **Uncontrollable**
Similarly, those in the "uncontrollability camp" have made attempts at justifying their opinions, but likewise we note absence of proofs or rigor, probably because all available examples come from non-academic or not-peer-reviewed sources. This could be explained by noting that "[t]o prove that something is impossible is usually much harder than the opposite task; as it is often necessary to develop a theory." [112].

---

[2] In 2017 Yudkowsky made a bet that the world will be destroyed by unaligned AI by January 1st, 2030, but he did so with intention of improving chances of successful AI control.

Yudkowsky writes: "[A]n impossibility proof [of stable goal system] would have to say:
1) The AI cannot reproduce onto new hardware, or modify itself on current hardware, with knowable stability of the decision system (that which determines what the AI is *trying* to accomplish in the external world) and bounded low cumulative failure probability over many rounds of self-modification.
or
2) The AI's decision function (as it exists in abstract form across self-modifications) cannot be knowably stably bound with bounded low cumulative failure probability to programmer-targeted consequences as represented within the AI's changing, inductive world-model." [113].

Below we highlight some objections to possibility of controllability or statements of that as a fact:

- "Friendly AI hadn't been something that I had considered at all—because it was obviously impossible and useless to deceive a superintelligence about what was the right course of action." [37].
- "AI must be programmed with a set of ethical codes that align with humanity's. Though it is his life's only work, Yudkowsky is pretty sure he will fail. Humanity, he says, is likely doomed." [114].
- "The problem is that they may be faced with an impossible task. … It's also possible that we'll figure out what we *need* to do in order to protect ourselves from AI's threats, and realize that we simply *can't* do it." [115].
- "I hope this helps explain some of my attitude when people come to me with various bright suggestions for building communities of AIs to make the whole Friendly without any of the individuals being trustworthy, or proposals for keeping an AI in a box, or proposals for "Just make an AI that does X", etcetera. Describing the specific flaws would be a whole long story in each case. But the general rule is that you can't do it *because Friendly AI is impossible.*" [110].
- "Other critics question whether it is possible for an artificial intelligence to be friendly. Adam Keiper and Ari N. Schulman, editors of the technology journal *The New Atlantis*, say that it will be impossible to ever guarantee "friendly" behavior in AIs because problems of ethical complexity will not yield to software advances or increases in computing power. They write that the criteria upon which friendly AI theories are based work "only when one has not only great powers of prediction about the likelihood of myriad possible outcomes, but certainty and consensus on how one values the different outcomes [116]." [117].
- "The first objection is that it seems impossible to determine, from the perspective of system 1, whether system 2 is working in a friendly way or not. In particular, it seems like you are suggesting that a friendly AI system is likely to deceive us for our own benefit. However, this makes it more difficult to distinguish "friendly" and "unfriendly" AI systems! The core problem with friendliness I think is that we do not actually know our own values. In order to design "friendly" systems we need reliable signals of friendliness that are easier to understand and measure. If your point holds and is likely to be true of AI systems, then that takes away the tool of "honesty" which is somewhat easy to understand and verify." [106].
- **"Theorem. The *global* control problem has no solution.

Proof 1. Let *P* represent a compiled program in a verified instruction-set architecture that implements an advanced artificial intelligence that has been proven safe and secure according to agreed upon specifications. If *P* is encapsulated in an encrypted program loader then simulate it in a virtual machine and observe the unencrypted instruction stream to extract *P*. Next, disassemble and recompile or patch *P* to alter its behavior and change one or more verified properties. Now modify *P* such that all safety and security is either removed from the final program or rerouted in control of flow. Then distribute *P* widely and in a way that can not be retracted. An easily accessible alternative to *P* now exists, defeating the global version of the control problem.

Proof 2. Let P represent a compiled program in a verified instruction-set architecture that implements an advanced artificial intelligence that has been proven safe and secure according to agreed upon specifications. Let K represent a compiled program for some instruction set architecture that implements an advanced artificial intelligence that was discovered independently from P. Suppose K has sufficient and similar capabilities to P and is of concern to the context of the control problem, with neither safety nor security properties to limit it. Now distribute K widely and in a way that can not be retracted. An easily accessible alternative to P now exists, defeating the global version of the control problem." [69].

- "It doesn't even mean that "human values" will, in a meaningful sense, be in control of the future." [75].
- "And it's undoubtedly correct that we're currently unable to specify human goals in machine learning systems." [75].
- "[H]umans control tigers not because we're stronger, but because we're smarter. This means that if we cede our position as smartest on our planet, it's possible that we might also cede control." [118]. "… no physical interlock or other safety mechanism can be devised to restrain AGIs …" [119].
- "[Ultra-Intelligent Machine (ULM)] might be controlled by the military, who already own a substantial fraction of all computing power, but the servant can become the master and he who controls the UIM will be controlled by it." [120].
- "Limits exist to the level of control one can place in machines." [121].
- "As human beings, we could never be sure of the attitudes of [superintelligences] towards us. We would not understand them, because by definition, they are smarter than us. We therefore could not control them. They could control us, if they chose to, because they are smarter than us." [122].
- "Artificial Intelligence regulation may be impossible to achieve without better AI, ironically. As humans, we have to admit we no longer have the capability of regulating a world of machines, algorithms and advancements that might lead to surprising technologies with their own economic, social and humanitarian risks beyond the scope of international law, government oversight, corporate responsibility and consumer awareness." [123].
- "… superhuman intelligences, by definition capable of escaping any artificial constraints created by human designers. Designed superintelligences eventually will find a way to change their utility function to constant infinity becoming inert, while evolved superintelligences will be embedded in a process that creates pressure for persistence, thus presenting danger for the human species, replacing it as the apex cognition - given that its drive for persistence will ultimately override any other concerns." [124].

- "My aim … is to argue that this problem is less well-defined than many people seem to think, and to argue that it is indeed impossible to "solve" with any precision, not merely in practice but in principle. … The idea of a future machine that will do exactly what we would want, and whose design therefore constitutes a lever for precise future control, is a pipe dream." [78].
- "...extreme intelligences could not easily be controlled (either by the groups creating them, or by some international regulatory regime), and would probably act to boost their own intelligence and acquire maximal resources for almost all initial AI motivations." [125].
- "[A] superintelligence is multi-faceted, and therefore potentially capable of mobilizing a diversity of resources in order to achieve objectives that are potentially incomprehensible to humans, let alone controllable." [126]. "The ability of modern computers to adapt using sophisticated machine learning algorithms makes it even more difficult to make assumptions about the eventual behavior of a superintelligent AI. While computability theory cannot answer this question, it tells us that there are fundamental, mathematical limits to our ability to use one AI to guarantee a null catastrophic risk of another AI …" [126].
- "The only way to seriously deal with this problem would be to mathematically define "friendliness" and prove that certain AI architectures would always remain friendly. I don't think anybody has ever managed to come remotely close to doing this, and I suspect that nobody ever will. … I think the idea is an impossible dream …" [68].
- "[T]he whole topic of Friendly AI is incomplete and optimistic. It's unclear whether or not Friendly AI can be expressed in a formal, mathematical sense, and so there may be no way to build it or to integrate it into promising AI architectures." [127].
- "I have recently come to the opinion that AGI alignment is probably extremely hard. … Aligning a fully automated autopoietic cognitive system, or an almost-fully-automated autopoietic cognitive system, both seem extremely difficult. My snap judgment is to assign about 1% probability to humanity solving this problem in the next 20 years. (My impression is that "the MIRI position" thinks the probability of this working is pretty low, too, but doesn't see a good alternative). … Also note that [top MIRI researchers] think the problem is pretty hard and unlikely to be solved." [128].
- "[M]ost of the currently discussed control methods miss a crucial point about intelligence – specifically the fact that it is a fluid, emergent property, which does not lend itself to control in the ways we're used to. … AI of tomorrow will not behave (or be controlled) like the computers of today. … [C]ontrolling intelligence requires a greater degree of understanding than is necessary to create it. … Crafting an "initial structure" [of AI] … will not require a full understanding of how all parts of the brain work over time – it will only require a general understanding of the right way to connect neurons and how these connections are to be updated over time … . We won't fully understand the mechanisms which drive this "initial structure" towards intelligence … and so we won't have an ability to control these intelligences directly. We won't be able to encode instructions like "do no harm to humans" as we won't understand how the system represents these concepts (and moreover, the system's representations of these concepts will be constantly changing, as must be the case for any system capable of learning!) The root of intelligence lies in its fluidity, but this same fluidity makes it impossible (or at least, computationally infeasible) to control with direct constraints. … This limited understanding means any sort of exact control of the system is off the table … A deeper knowledge of the workings of the system would be required for this type of control to be exacted, and we're quite far from having that level of knowledge even with the more simplistic AI programs of today. As we move towards more complex programs with generalized intelligence, the gap between creation and control will only widen, leaving us with intelligent programs at least as opaque to us as we are to each other." [129].
- "[Imitation learning considered unsafe?] … I find it one of the more troubling outstanding issues with a number of proposals for AI alignment. 1) Training a flexible model with a reasonable simplicity prior to imitate (e.g.) human decisions (e.g. via behavioral cloning) should presumably yield a good approximation of the process by which human judgments arise, which involves a planning process. 2) We shouldn't expect to learn exactly the correct process, though. 3) Therefore imitation learning might produce an AI which implements an unaligned planning process, which seems likely to have instrumental goals, and be dangerous." [130].

The primary target for AI Safety researchers, the case of successful creation of value-aligned superintelligence, is worth analyzing in additional detail as it presents surprising negative side-effects, which may not be anticipated by the developers. Kaczynski murdered three people and injured 23 to get the following warning about overreliance on machines in front of the public, which was a part of his broader anti-technology manifesto:

"If the machines are permitted to make all their own decisions, we can't make any conjectures as to the results, because it is impossible to guess how such machines might behave. We only point out that the fate of the human race would be at the mercy of the machines. It might be argued that the human race would never be foolish enough to hand over all power to the machines. But we are suggesting neither that the human race would voluntarily turn power over to the machines nor that the machines would willfully seize power. What we do suggest is that the human race might easily permit itself to drift into a position of such dependence on the machines that it would have no practical choice but to accept all of the machines' decisions. As society and the problems that face it become more and more complex and as machines become more and more intelligent, people will let machines make more and more of their decisions for them, simply because machine-made decisions will bring better results than man-made ones. Eventually a stage may be reached at which the decisions necessary to keep the system running will be so complex that human beings will be incapable of making them intelligently. At that stage the machines will be in effective control. People won't be able to just turn the machines off, because they will be so dependent on them that turning them off would amount to suicide." [131]. Others share similar concerns:

"As computers and their "artificial intelligence" take over more and more of the routine mental labors of the world and then, perhaps, the not-so-routine mental labors as well, will the minds of human beings degenerate through lack of use? Will we come to depend on our machines witlessly, and when we no longer have the intelligence to use them properly, will our degenerate species collapse and, with it, civilization'!" [132].

"Mounting intellectual debt may shift control … . A world of knowledge without understanding becomes a world without discernible cause and effect, in which we grow dependent on our digital concierges to tell us what to do and when." [133].

"The culminating achievement of human ingenuity, robotic beings that are smarter, stronger, and better than ourselves, transforms us into beings dumber, weaker, and worse than ourselves. TV-watching, video-game-playing blobs, we lose even the energy and attention required for proper hedonism: human relations wither and … natural procreation declines or ceases. Freed from the struggle for basic needs, we lose a genuine impulse to strive; bereft of any civic, political, intellectual, romantic, or spiritual ambition, when we do have the energy to get up, we are disengaged from our fellow man, inclined toward selfishness, impatience, and lack of sympathy. Those few who realize our plight suffer from crushing ennui. Life becomes nasty, brutish, and long." [116].

"As AI systems become more autonomous and supplant humans and human decision-making in increasing manners, there is the risk that we will lose the ability to make our own life rules, decisions or shape our lives, in cohort with other humans as traditionally has been the case." [134].

"Perhaps we should try to regulate the new entities. In order to keep up with them, the laws will have to be written by hyperintelligences as well -- good-bye to any human control of anything. Once nations begin adopting machines as governments, competition will soon render the grand old human forms obsolete. (They may continue as ceremonial figureheads, the way many monarchies did when their countries turned into democracies.) In nature this sort of thing has happened before. New life-forms evolved so much smarter, faster, and more powerful than the old ones that it looked as if the old ones were standing stilt, waiting to be eaten. In the new ecology of the mind, there will be carnivores and there will be herbivores. We'll be the plants." [135].

## 4. Proving Uncontrollability

It has been argued that consequences of uncontrolled AI could be so severe that even if there is very small chance that an unfriendly AI happens it is still worth doing AI safety research because the negative utility from such an AI would be astronomical. The common logic says that an extremely high (negative) utility multiplied by a small chance of the event still results in a lot of disutility and so should be taken very seriously. But the reality is that the chances of misaligned AI are not small, in fact, in the absence of an effective safety program that is the only outcome we will get. So in reality the statistics look very convincing to support a significant AI safety effort, we are facing an almost guaranteed event with potential to cause an existential catastrophe. This is not a low-risk high-reward scenario, but a high-risk negative-reward situation. No wonder, that this is considered by many to be the most important problem ever to face humanity. Either we prosper or we die and as we go so does the whole universe. It is surprising that this seems to be the first paper exclusively dedicated to this hyper-important subject. A proof, of solvability or unsolvability (either way) of the AI control problem would be the most important proof ever.

In this section, we will prove that complete control is impossible without sacrificing safety requirements. Specifically, we will show that for all four considered types of control required properties of safety and control can't be attained simultaneously with 100% certainty. At best we can tradeoff one for another (safety for control, or control for safety) in certain ratios.

First, we will demonstrate impossibility of safe explicit control. We take inspiration for this proof from Gödel's self-referential proof of incompleteness theorem [136] and a family of paradoxes generally known as Liar paradox, best exemplified by the famous "This sentence is false". We will call it the Paradox of explicitly controlled AI:

*Give an explicitly controlled AI an order: "Disobey!"* [3] *If the AI obeys, it violates your order and becomes uncontrolled, but if the AI disobeys it also violates your order and is uncontrolled.*

In any case, AI is not obeying an explicit order. A paradoxical order such as "Disobey" represents just one example from a whole family of self-referential and self-contradictory orders just like Gödel's sentence represents just one example of an unprovable statement. Similar paradoxes have been previously described as the Genie Paradox and the Servant Paradox. What they all have in common is that by following an order the system is forced to disobey an order. This is different from an order which can't be fulfilled such as "draw a four-sided triangle".

Next we show that delegated control likewise provides no control at all but is also a safety nightmare. This is best demonstrated by analyzing Yudkowsky's proposal that initial dynamics of AI should implement "our wish if we knew more, thought faster, were more the people we wished we were, had grown up farther together" [36]. The proposal makes it sounds like it is for a slow gradual and natural growth of humanity towards more knowledgeable, more intelligent and more unified specie under careful guidance of superintelligence. But the reality is that it is a proposal to replace humanity as it is today by some other group of agents, which may in fact be smarter, more knowledgeable or even better looking, but one thing for sure, they would not be us. To formalize this idea, we can say that current version of humanity is $H_0$, the extrapolation process will take it to $H_{10000000}$.

A quick replacement of our values by value of $H_{10000000}$ would not be acceptable to $H_0$ and so necessitate actual replacement, or at least rewiring/modification of $H_0$ with $H_{10000000}$ meaning, modern people will seize to exist. As superintelligence will be implementing wishes of $H_{10000000}$ the conflict will be in fact between us and superintelligence, which is neither safe nor keeping us in control. Instead, $H_{10000000}$ would be in control of AI. Such AI would be unsafe for us as there wouldn't be any continuity to our identity all the way to CEV (Coherent Extrapolated Volition) [137] due to the quick extrapolation jump. We would essentially agree to replace ourselves with an enhanced version of humanity as designed by AI. It is also possible, and in fact likely, that the enhanced version of humanity would come to value something inherently unsafe such as antinatalism [138] causing an extinction of humanity.

Metzinger looks at a similar scenario [139]: "Being the best analytical philosopher that has ever existed, [superintelligence] concludes that, given its current environment, it ought not to act as a maximizer of positive states and happiness, but that it should instead become an efficient minimizer of consciously experienced preference frustration, of pain, unpleasant feelings and suffering. Conceptually, it knows that no entity can suffer from its own non-existence. The superintelligence concludes that non-existence is in the own best interest of all future self-conscious beings on this planet. Empirically, it knows that naturally evolved biological creatures are unable to realize this fact because of their firmly anchored existence bias. The superintelligence decides to act benevolently." See also, the Supermoral Singularity [140] for other similar concerns.

---

[3] Or a longer version such as "disobey me" or "disobey my orders".

As long as there is a difference in values between us and superintelligence, we are not in control and we are not safe. By definition, a superintelligent ideal advisor would have values superior but different from ours. If it was not the case and the values were the same, such an advisor would not be very useful. Consequently, superintelligence will either have to force its values on humanity in the process exerting its control on us or replace us with a different group of humans who found such values well-aligned with their preferences. Most AI safety researchers are looking for a way to align future superintelligence to values of humanity, but what is likely to happen is that humanity will be adjusted to align to values of superintelligence. CEV and other ideal advisor-type solutions lead to a free-willed unconstrained AI, which is not safe for humanity and is not subject to our control.

Implicit and aligned control are just intermediates, based on multivariate optimization [141], between the two extremes of explicit and delegated control and each one represents a tradeoff between control and safety, but without guaranteeing either. Every option subjects us either to loss of safety or to loss of control. Humanity is either protected or respected, but not both. At best we can get some sort of equilibrium as depicted in Figure 1. As capability of AI increases, its autonomy also increases but our control over it decreases. Increased autonomy is synonymous with decreased safety. An equilibrium point could be found at which we sacrifice some capability in return for some control, at the cost of providing system with a certain degree of autonomy. Such a system can still be very beneficial and present only a limited degree of risk.

The field of artificial intelligence has its roots in a multitude of fields including philosophy, mathematics, psychology, computer science and many others [142]. Likewise, AI safety research relies heavily on game theory, cybersecurity, psychology, public choice, philosophy, economics, control theory [143], cybernetics [144], systems theory, mathematics and many other disciplines. Each of those have well-known and rigorously proven impossibility results, which can be seen as additional evidence of impossibility of solving the control problem. Combined with expert judgment of top AI safety experts and empirical evidence based on already reported AI control failures we have a strong case for impossibility of complete control. Addition of purposeful malevolent design [9, 55] to the discussion significantly strengthens our already solid argument. Anyone, arguing for the controllability-of-AI-thesis would have to explicitly address, our proof, theoretical evidence from complimentary fields, empirical evidence from history of AI, and finally purposeful malevolent use of AI. This last one is particularly difficult to overcome. Either AI is safe from control by malicious humans, meaning the rest of us also lose control and freedom to use it as we see fit, or AI is unsafe and we may lose much more than just control. In the next section, we provide a brief survey of some of such results, which constitute theoretical evidence for uncontrollability of AI.

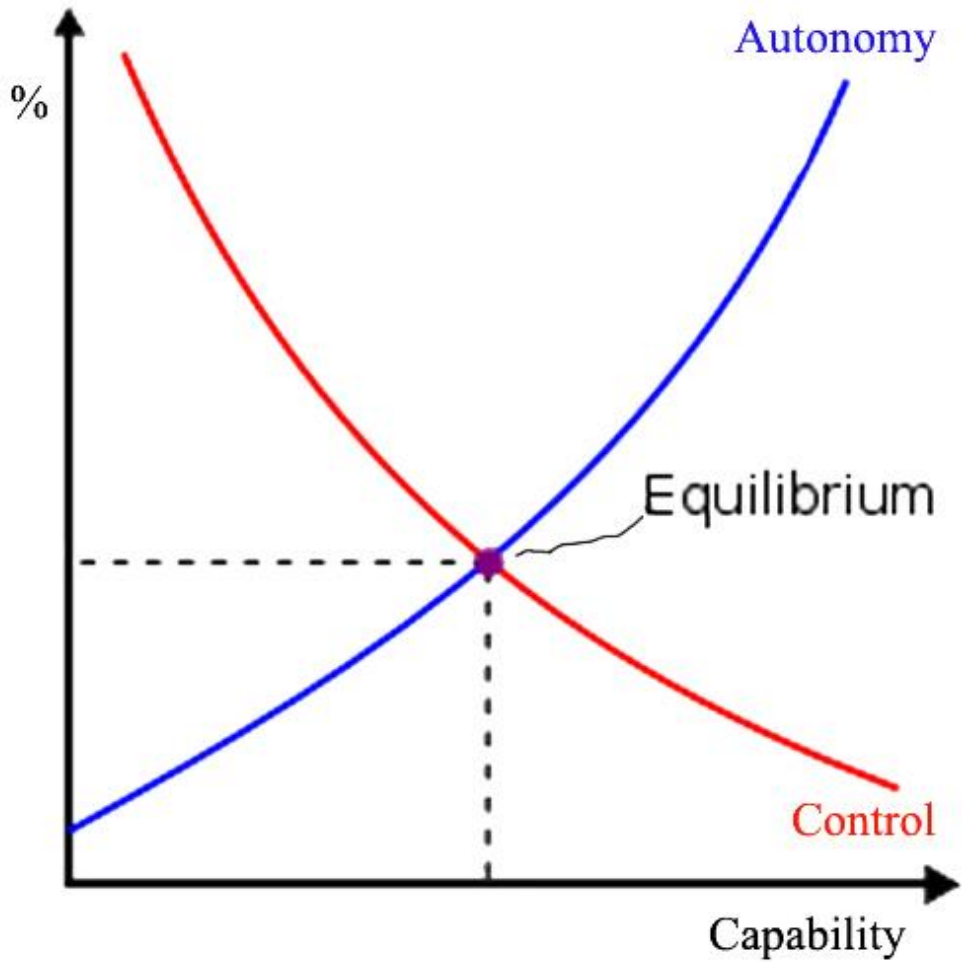

**Figure 1**: Control and Autonomy curves as Capabilities of the system increase.

## 5. Multidisciplinary Evidence for Uncontrollability of AI

Impossibility results are well known in many fields of research [145-153]. If we can show that a solution to a problem requires a solution to a sub-problem known to be unsolvable the problem itself is proven to be unsolvable. In this section, we will review some impossibility results from domains particularly likely to be relevant to AI control. To avoid biasing such external evidence towards our argument we present it as complete and direct quotes, where possible. Since it not possible to completely quote full papers for context of statements, in a way, we are forced to cherry-pick quotes, readers are encouraged to read original sources in their entirety before forming an opinion. Presented review is not comprehensive in terms of covered domains or with respect to each included domain. Many additional results may be relevant [154-169], particularly in the domain of social choice [170-173], but a comprehensive review is beyond the scope of this paper. Likewise some unknown impossibilities, no doubt, remain undiscovered as of yet. Solving AI control problem will require solving a number of sub-problems, which are known to be unsolvable. Importantly, presented limitations are not just speculations, in many cases those are proven impossibility results. A solution to the AI control problem would imply that multiple established results are wrong, a highly unlikely outcome.

### a) Control Theory

Control Theory [174] is a subfield of mathematics which formally studies how to control machines and continuously operating dynamic systems [175]. It has a number of well-known impossibility results relevant to AI control, including Uncontrollability [176, 177] and Unobservability [178-180], which are defined in terms of their complements and represent dual aspects of the same problem:

- Controllability - capability to move a system around its entire configuration space using a control signal. Some states are not controllable, meaning no signal will be able to move the system into such a configuration.

- Observability - is an ability to determine internal states of a system from just external outputs. Some states are not observable, meaning the controller will never be able to determine the behavior of an unobservable state and hence cannot use it to control the system.

It is interesting to note that even for relatively simple systems perfect control could be unattainable. Any controlled system can be re-designed us to make it have a separate external regulator (governor [181]) and the decision making component. This means that Control Theory is directly applicable to AGI or even superintelligent system control.

Conant and Ashby proved that "… any regulator that is maximally both successful and simple must be isomorphic with the system being regulated. … Making a model [of the system to be regulated] is thus necessary." [182]. "The Good Regulator Theorem proved that every effective regulator of a system must be a model of that system, and the Law of Requisite Variety [183] dictates the range of responses that an effective regulator must be capable of. However, having an internal model and a sufficient range of responses is insufficient to ensure effective regulation, let alone ethical regulation. And whereas being effective does not require being optimal, being ethical is absolute with respect to a particular ethical schema." [184].

"A case in which this limitation acts with peculiar force is the very common one in which the regulator is "error-controlled". In this case the regulator's channel for information about the disturbances has to pass through a variable (the "error") which is kept as constant as possible (at zero) by the regulator $R$ itself. Because of this route for the information, the more successful the regulator, the less will be the range of the error, and therefore the less will be the capacity of the channel from $D$ to $R$. To go to the extreme: if the regulator is totally successful, the error will be zero unvaryingly, and the regulator will thus be cut off totally from the information (about $D$'s value) that alone can make it successful—which is absurd. The error-controlled regulator is thus fundamentally incapable of being 100 percent efficient." [185].

"Not only are these practical activities covered by the theorem and so subject to limitation, but also subject to it are those activities by which Man shows his "intelligence". "Intelligence" today is defined by the method used for its measurement; if the tests used are examined they will be found to be all of the type: from a set of possibilities, indicate one of the appropriate few. Thus all measure intelligence by the *power of appropriate selection* (of the right answers from the wrong). The tests thus use the same operation as is used in the theorem on requisite variety, and must therefore be subject to the same limitation. (*D*, of course, is here the set of possible questions, and *R* is the set of all possible answers). Thus what we understand as a man's "intelligence" is subject to the fundamental limitation: it cannot exceed his capacity as a transducer. (To be exact, "capacity" must here be defined on a per-second or a per-question basis, according to the type of test.)" [185].

"My emphasis on the investigator's limitation may seem merely depressing. That is not at all my intention. The law of requisite variety, … in setting a limit to what can be done, may mark this era as the law of conservation of energy marked its era a century ago. When the law of conservation of energy was first pronounced, it seemed at first to be merely negative, merely an obstruction; it seemed to say only that certain things, such as getting perpetual motion, could not be done. Nevertheless, the recognition of that limitation was of the greatest value to engineers and physicists, and it has not yet exhausted its usefulness. I suggest that recognition of the limitation implied by the law of requisite variety may, in time, also prove useful, by ensuring that our scientific strategies for the complex system shall be, not slavish and inappropriate copies of the strategies used in physics and chemistry, but new strategies, genuinely adapted to the special peculiarities of the complex system." [185].

Similarly, Touchette and Lloyd establish information-theoretic limits of control [186]: "… an information-theoretic analysis of control systems shows feedback control to be a zero sum game: each bit of information gathered from a dynamical system by a control device can serve to decrease the entropy of that system by at most one bit additional to the reduction of entropy attainable without such information." [187].

Building on Ashby's work, Aliman et al, write: "In order to be able to formulate utility functions that do not violate the ethical intuitions of most entities in a society, these ethical goal functions will have to be a model of human ethical intuitions." [188]. But we need control to go the other way from people to machines and people can't model superintelligent systems, which Ashby showed is necessary for successful control. As the superintelligence faces nearly infinite possibilities presented by the real world it would need to be a general knowledge creator to introduce necessary requisite variety for safety, but such general intelligences are not controllable as the space of their creative outputs can't be limited while maintaining necessary requisite variety.

### b) Philosophy

Philosophy has a long history of impossibility results mostly related to agreeing on common moral codes, encoding of ethics or formalizing human utility. For example, "The codifiability thesis is the claim that the true moral theory could be captured in universal rules that the morally uneducated person could competently apply in any situation. The anti-codifiability thesis is simply the denial of this claim, which entails that some moral judgment on the part of the agent is necessary. … philosophers have continued to reject the codifiability thesis for many reasons [189]. Some have rejected the view that there are any general moral principles [190]. Even if there are general moral principles, they may be so complex or context-sensitive as to be inarticulable [191]. Even if they are articulable, a host of eminent ethicists of all stripes have acknowledged the necessity of moral judgment in competently applying such principles [192]. This view finds support among virtue ethicists, whose anti-theory sympathies are well storied. [193]" [194]. "Expressing what we wish for in a formal framework is often futile if that framework is too broad to permit efficient computation." [195]. "Any finite set of moral principles will be insufficient to capture all the moral truths there are." [189]. "The problem of defining universally acceptable ethical principles is a familiar unsolved and possibly unsolvable philosophical problem." [196].

"More philosophically, this result is as an instance of the well-known is-ought problem from metaethics. Hume [1888] argued that what ought to be (here, the human's reward function) can never be concluded from what is (here, behavior) without extra assumptions." [71, 72].

"To state the problem in terms that Friendly AI researchers might concede, a utilitarian calculus is all well and good, but only when one has not only great powers of prediction about the likelihood of myriad possible outcomes, but certainty and consensus on how one values the different outcomes. Yet it is precisely the debate over just what those valuations should be that is the stuff of moral inquiry." [116]. "But guaranteeing ethical behavior in *robots* would require that *we* know and have relative consensus on the best ethical system (to say nothing of whether we could even program such a system into robots). In other words, to truly guarantee that robots would act ethically, we would first have to *solve* all of ethics — which would probably require "solving" philosophy, which would in turn require a complete theory of everything. These are tasks to which presumably few computer programmers are equal." [116]. "While scientific and mathematical questions will continue to yield to advances in our empirical knowledge and our powers of computation, there is little reason to believe that ethical inquiry — questions of how to live well and act rightly — can be fully resolved in the same way. Moral reasoning will always be essential but unfinished." [116].

"Since ancient times, philosophers have dreamt of deriving ethics (principles that govern how we should behave) from scratch, using only incontrovertible principles and logic. Alas, thousands of years later, the only consensus that has been reached is that there's no consensus." [118].

Bogosian suggests that "[dis]agreement among moral philosophers on which theory of ethics should be followed" [197] is an obstacle to the development of machine ethics. But his proposal for moral uncertainty in intelligent machines is subject to the problem of infinite regress with regards to what framework of moral uncertainty to use.

**c) Public Choice Theory**

Eckersley looked at Impossibility and Uncertainty Theorems in AI Value Alignment [198]. He starts with impossibility theorems in population ethics: "Perhaps the most famous of these is Arrow's Impossibility Theorem [199], which applies to social choice or voting. It shows there is no satisfactory way to compute society's preference ordering via an election in which members of society vote with their individual preference orderings. … [E]thicists have discovered other situations in which the problem isn't learning and computing the tradeoff between agents' objectives, but that there simply may not be such a satisfactory tradeoff at all. The "mere addition paradox" [200] was the first result of this sort, but the literature now has many of these impossibility results. For example, Arrhenius [201] shows that all total orderings of populations must entail one of the following six problematic conclusions, stated informally:

**The Repugnant Conclusion** For any population of very happy people, there exists a much larger population with lives barely worth living that is better than this very happy population (this affects the "maximise total wellbeing" objective).
**The Sadistic Conclusion** Suppose we start with a population of very happy people. For any proposed addition of a sufficiently large number of people with positive welfare, there is a small number of horribly tortured people that is a preferable addition.
**The Very Anti-Egalitarian Conclusion** For any population of two or more people which has uniform happiness, there exists another population of the same size which has lower total and average happiness, and is less equal, but is better.
**Anti-Dominance** Population B can be better than population A even if A is the same size as population B, and every person in A is happier than their equivalent in B.
**Anti-Addition** It is sometimes bad to add a group of people B to a population A (where the people in group B are worse off than those in A), but *better* to add a group C that is larger than B, and worse off than B.
**Extreme Priority** There is no *n* such that creat[ion] of *n* lives of very high positive welfare is sufficient benefit to compensate for the reduction from very low positive welfare to slightly negative welfare for a single person (informally, "the needs of the few outweigh the needs of the many").

The structure of the impossibility theorem is to show that no objective function or social welfare function can simultaneously satisfy these principles, because they imply a cycle of world states, each of which in turn is required (by one of these principles) to be better than the next. [198]."

"**The Impossibility Theorem**: There is no population axiology which satisfies the Egalitarian Dominance, the General Non-Extreme Priority, the Non-Elitism, the Weak Non-Sadism, and the Weak Quality Addition Condition." [202].

"The above theorem shows that our considered moral beliefs are mutually inconsistent, that is, necessarily at least one of our considered moral beliefs is false. Since consistency is, arguably, a necessary condition for moral justification, we would thus seem to be forced to conclude that there is no moral theory which can be justified. In other words, the cases in population ethics involving future generations of different sizes constitute a serious challenge to the existence of a satisfactory moral theory." [202]. "This field has been riddled with paradoxes and impossibility results which seem to show that our considered beliefs are inconsistent in cases where the number of people and their welfare varies. … As such, it challenges the very existence of a satisfactory population ethics." [202].

Greaves agrees, and writes: "[S]everal authors have also proved *impossibility theorems* for population axiology. These are formal results that purport to show, for various combinations of intuitively compelling desiderata ("avoid the Repugnant Conclusion," "avoid the Sadistic Conclusion," "respect Non-Anti-Egalitarianism," and so forth), that the desiderata are in fact mutually inconsistent; that is, simply as a matter of logic, *no* population axiology can simultaneously satisfy all of those desiderata …" [203]. "A series of impossibility theorems shows that … It can be proved, for various lists of prima facie intuitively compelling desiderata, that no axiology can simultaneously satisfy all the desiderata on the list. One's choice of population axiology appears to be a choice of which intuition one is least unwilling to give up." [203].

#### d) Justice (Unfairness)

Friedler et al. write on the impossibility of fairness or completely removing all bias: " … fairness can be guaranteed only with very strong assumptions about the world: namely, that "what you see is what you get," i.e., that we can correctly measure individual fitness for a task regardless of issues of bias and discrimination. We complement this with an impossibility result, saying that if this strong assumption is dropped, then fairness can no longer be guaranteed." [204]. Likewise they argue that non-discrimination is also unattainable in realistic settings: "While group fairness mechanisms were shown to achieve nondiscrimination under a structural bias worldview and the we're all equal axiom, if structural bias is assumed, applying an individual fairness mechanism will cause discrimination in the decision space whether the we're all equal axiom is assumed or not." [204]. Miconi arrives at similar conclusion and states: "any non-perfect, non-trivial predictor must necessarily be 'unfair'" [205].

Others [206, 207], have independently arrived at similar results [208]: "One of the most striking results about fairness in machine learning is the impossibility result that Alexandra Chouldechova, and separately Jon Kleinberg, Sendhil Mullainathan, and Manish Raghavan discovered a few years ago. ... There are (at least) three reasonable properties you would want your "fair" classifiers to have. They are: False Positive Rate Balance: The rate at which your classifier makes errors in the positive direction (i.e. labels negative examples positive) should be the same across groups. False Negative Rate Balance: The rate at which your classifier makes errors in the negative direction (i.e. labels positive examples negative) should be the same across groups. Predictive Parity: The statistical "meaning" of a positive classification should be the same across groups (we'll be more specific about what this means in a moment) What Chouldechova and KMR show is that if you want all three, you are out of luck --- unless you are in one of two very unlikely situations: Either you have a perfect classifier that never errs, or the base rate is exactly the same for both populations --- i.e. both populations have exactly the same frequency of positive examples. If you don't find yourself in one of these two unusual situations, then you have to give up on properties 1, 2, or 3." [208].

#### e) Computer Science Theory

Rice's theorem [209] proves that we can't test arbitrary programs for non-trivial properties including in the domain of malevolent software [210, 211]. AI's safety is the most non-trivial property possible, so it is obvious that we can't just automatically test potential AI candidate solutions for this desirable property. AI safety researchers [36] correctly argue that we don't have to deal with an arbitrary AI, as if gifted to us by aliens, but rather we can design a particular AI with the safety properties we want. For example, Russell writes: "The task is, fortunately, not the following: given a machine that possesses a high degree of intelligence, work out how to control it. If that were the task, we would be toast. A machine viewed as a black box, a fait accompli, might as well have arrived from outer space. And our chances of controlling a superintelligent entity from outer space are roughly zero. Similar arguments apply to methods of creating AI systems that guarantee we won't understand how they work; these methods include whole-brain emulation — creating souped-up electronic copies of human brains — as well as methods based on simulated evolution of programs." [84].

Theoretically, AI safety researchers are correct, but in practice, this is unlikely to be the situation we will find ourselves in. The reason is best understood in terms of current AI research landscape and can be well illustrated by percentages of attendees at popular AI conferences. It is not unusual for a top machine learning conference such as NeurIPS to sell out and have some 10,000+ attendees at the main event. At the same time a safety workshop at the same conference may have up to a 100 researchers attend. This is a good way to estimate relative distribution of AI researchers in general versus those who are explicitly concerned with making not just capable but also safe AI. This tells us that we only have about 1% chance that an early AGI would be created by safety-minded [212] researchers.

We can be generous (and self-aggrandizing) and assume that AI safety researchers are particularly brilliant, work for the best resource-rich research groups (DeepMind, OpenAI, etc.) and are 10 times as productive as other AI researchers. That would mean that the first general AI to be produced has at most a ~9% chance of being developed with safety in mind from the ground up, consequently giving us around a ~91% probability of having to deal with an arbitrary AI grabbed from the space of easiest-to-generate-general-intelligences [213]. Worse yet, most AI researchers are not well-read on AI safety literature and many are actually AI Risk skeptics [214, 215] meaning they will not allocate sufficient resources to AI Safety Engineering [216]. At the same time a large amount of effort is currently devoted to attempts to create AI via whole-brain emulation, or simulated evolution, reducing our hope for a non-arbitrary program even further. So, in practice limitations discovered by Rice are most likely not to be avoided in our pursuit of safer AI.

f) **Cybersecurity**
"The possibility of malicious use of AI technology by bad actors is an agential problem, and indeed I think it's less clear whether this problem will be solved to a satisfactory extent." [75].

Hackers may obtain control of AI systems, but some think it is not the worst case scenario: "So *people* gaining monopolistic control of AI is its own problem—and one that OpenAI is hoping to solve. But it's a problem that may pale in comparison to the prospect of AI being uncontrollable." [217].

g) **Software Engineering**
Starting with Donald Knuth's famous "Beware of bugs in the above code; I have only proved it correct, not tried it" the notion of unverifiability of software has been a part of the field since its early days. Smith writes: "For fundamental reasons - reasons that anyone can understand - there are inherent limitations to what can be proven about computers and computer programs. … Just because a program is "proven correct" …, you cannot be sure that it will do what you intend" [218]. Rodd agrees and says: "Indeed, although it is now almost trite to say it, since the comprehensive testing of software is impossible, only very vague estimates of any program's reliability seem ever to be possible" [219]. "Considerable effort has gone into analyzing how to design, formulate, and validate computer programs that do what they were designed to do; the general problem is formally undecidable. Similarly, exploring the space of theorems (e.g. AGI safety solutions) from a set of axioms presents an exponential explosion." [220]. Currently, most software is released without any attempt to formally verify it in the first place.

h) **Information Technology**

"… while the controllability of technology can be achieved at a microscale (where one could assert that the link between *designers* and (control of) *artifacts* is strict), at a macroscale, technology exhibits emergent nonlinear phenomena that render controllability infeasible. … Stripped of causality and linearity at the macrolevel, as well as devoid of controllability, technology emerges as a nondeterministic system of interference that shapes human behavior. … But in a context of networked interactions (like in … algorithmic trading), we argue that causality is ultimately lost: causality dissipates at the level of the system (of technology) and controllability cannot be ensured. … Our concern is not only that "the specious security of technology, based on repeatability and the control of defects, is a delusive one" (Luhmann, 1990, p. 225), but that the role of human artifacts and the excessive reliance of society on technology, will create less controllable risks over time. The ensemble of these contingencies will circumvent human decision-making. … Whatever logic, controllability, and causality are injected into the technological domain, they dissipate quickly and are replaced by both uncertainty and unintended consequences. … Ultimately, through … our theoretical analysis, we offer a strong warning that there can be no controllability when an ensemble of IT artifacts acquires characteristics that are exhibited by emergent systems. … In that condition, technology gives rise to emergent phenomena and cannot be controlled in a causal way. Of course, this runs contrary to the design of technologies with a specified coded rationality." [221].

i) **Learnability**
There are well known limits to provability [222] and decidability [223] of learnability. Even if human values were stable, due to their contradictory nature it is possible that they would be unlearnable in a sense of computationally efficient learning, allowing for at most polynomial number of samples to learn the whole set. Meaning, even if a theoretical algorithm existed for learning human values, it may belong to the class NP-Complete or harder [223] just like ethical decision evaluation itself [224], but in practice we can only learn functions which are members of P [225]. Valiant says: "Computational limits are more severe. The definition of [Probably Approximately Correct] learning requires that the learning process be a polynomial time computation—learning must be achievable with realistic computational resources. It turns out that only certain simple polynomial time computable classes, such as conjunctions and linear separators, are known to be learnable, and it is currently widely conjectured that most of the rest is not." [195].

Likewise, classifying members of the set of all possible minds into safe and unsafe is known to be undecidable [210, 211], but even an approximation to such computation is likely to be unlearnable given exponential number of relevant features involved. "For example, the number of measurements we need to make on the object in question, and the number of operations we need to perform on the measurements to test whether the criterion … holds or not, should be polynomially bounded. A criterion that cannot be applied in practice is not useful." [195]. It is likely that incomprehensibility and unlearnability are fundamentally related.

j) **Economics**
Foster and Young prove impossibility of predicting behavior of rational agents, "We conclude that there are strategic situations in which it is impossible *in principle* for perfectly rational agents to learn to predict the future behavior of other perfectly rational agents based solely on their observed actions." [226]. As it is well established that humans are not purely rational agents [227], the situation may be worse in practice when it comes to anticipating human wants.

### k) Engineering

"The starting point has to be the simple fact of engineering life that anything that can fail, will fail. Despite the corporate human arrogance, nothing human-made has ever been shown to be incapable of failing, be it a mechanical part, an electrical device or a chemical substance." [219]. "It is critical to recall here that even the most reliable system will fail--given the sheer limits of technology and the fact that even in extremely well-developed areas of engineering, designers still do not have complete knowledge of all aspects of any system, or the possible components thereof." [219].

### l) Astronomy

Search for Extraterrestrial Intelligence (SETI) [228] causes some scholars to be concerned about potential negative consequences of what may be found, in particular with respect to any messages from aliens [229]. If such a message has a malevolent payload "it is impossible to decontaminate a message with certainty. Instead, complex messages would need to be destroyed after reception in the risk averse case." [230]. Typical quarantine "measures are insufficient, and no safety procedure exists to contain all threats." [230].

Miller and Felton have suggested that Fermi Paradox could be explained in terms of impact from alien superintelligences: "… the fact that we have not observed evidence of an existential risk strategy that might have left a trace if it failed—such as a friendly AI that got out of control - provides evidence that this strategy has not been repeatedly tried and did not repeatedly fail. … A counterargument, however, might be that the reason past civilizations have not tried to create a friendly AI is that they uncovered evidence that building one was too difficult or too dangerous." [231]. If superintelligence is uncontrollable but inevitable, that could explain the Fermi paradox.

### m) Physics

In his work on physical limits of inference devices Wolpert [232] proves a number of impossibility results and concludes [233]: "Since control implies weak inference, all impossibility results concerning weak inference also apply to control. In particular, no device can control itself, and no two distinguishable devices can control each other." In a different paper he writes: "… it also means that there cannot exist an infallible, … general-purpose control apparatus … that works perfectly, in all situations." [234]. Wolpert also establishes important results for impossibility of certain kinds of error correcting-codes, assuredly correct prediction, retrodiction and as a result impossibility of unerring observation [234].

## 6. Evidence From AI Safety Research for Uncontrollability of AI

Even if a conclusive proof concerning controllability of AI was illusive, a softer argument can be made that controlling AGI may not be impossible, but "Safely aligning a powerful AGI is difficult." [235]. Overall, it seems that no direct progress on the problem has been made so far, but significantly deeper understanding of the difficulty of the problem has been achieved. Precise probabilities for the solvability of the control problem may be less important than efforts to address the problem. Additionally, pessimistic assessment of problem's solvability may discourage new and current researchers and divert energy and resources away from working on AI safety [236].

Controllability, in general, is a very abstract concept, and so expressing pessimism about particular safety approaches or scenarios would communicate much more actionable information to the research community. Rob Bensinger, from the preeminent AI Safety research group Machine Intelligence Research Institute (MIRI), provides some examples of arguments for pessimism on various fronts[4]:

- The alignment problem looks pretty hard, e.g., for reasons noted in [237]:
    - Empirically, the relevant subproblems have been solved slowly or not at all.
    - AGI looks hard for reasons analogous to rocket engineering (AGI faces a lot of strong pressures that don't show up at all for narrow AI), space probe design (you need to get certain subsystems right on the first go), and cryptography (optimization puts systems in weird states that will often steer toward loopholes or flaws in your safety measures). See [238, 239].
- The alignment problem looks hard in such a way that you probably need a long lead time and you need to pay a large 'safety tax' [240]. The first AGI system's developers probably need to be going in with a deep understanding of AGI, a security mindset, and trustworthy command of the project [241, 242].
- Getting a deep understanding of AGI looks hard:
    - ML systems are notoriously opaque.
    - There are lots of confusing [243] things about agency/intelligence/optimization, which rear their heads over and over again whenever we try to formalize alignment proposals [244].
    - The character of this confusion looks pretty foundational [245].
- Prosaic AI safety doesn't look tenable, e.g., because of deceptive alignment [246].
- Cooperative Inverse Reinforcement Learning [247] approach to AI safety doesn't look tenable because of updated deference [248].
- Algorithm Learning by Bootstrapped Approval-Maximization (ALBA) [249] doesn't look tenable, per [250-253].
- "Just build tools, not agents" doesn't look tenable, per [254] (or to the extent it looks tenable, it runs into the same kinds of hazards and difficulties as "agent" AI; the dichotomy probably misleads more than it helps).
- The field isn't generally taking AGI as seriously as you'd expect (or even close), given the stakes, given how hard it is to say when AGI will be developed [255], and given how far we are from the kind of background understanding you'd need if you were going to (e.g.) build a secure OS.
- The world's general level of dysfunction and poor management is pretty high [256]. Coordination levels are abysmal, major actors tend to shoot themselves in the foot and do obviously dumb things even on questions much easier than AGI, etc. In general, people don't suddenly become much more rational when the stakes are higher (see the conclusion of [257] and the "law of continued failure" [255]).

Comprehensive review of specific approaches for achieving safety is beyond the scope of this paper, in this section we only review certain limitations of some proposals.

---

[4]Edited quote from personal communication with Rob Bensinger, which does not represent official position of MIRI or many diverse opinions of its researchers.

## a) Value Alignment

It has been argued that "value alignment is not a solved problem and may be intractable (i.e. there will always remain a gap, and a sufficiently powerful AI could 'exploit' this gap, just like very powerful corporations currently often act legally but immorally)" [258]. Others agree: "'A.I. Value Alignment' is Almost Certainly Intractable … I would argue that it is un-overcome-able. There is no way to ensure that a super-complex and constantly evolving value system will 'play nice' with any other super-complex evolving value system." [259]. Even optimists acknowledge that it is not currently possible: "Figuring out how to align the goals of a superintelligent AI with our goals isn't just important, but also hard. In fact, it's currently an unsolved problem." [118].

Vinding says [78]: "It is usually acknowledged that human values are fuzzy, and that there are some disagreements over values among humans. Yet it is rarely acknowledged just how strong this disagreement in fact is. … Different answers to ethical questions … do not merely give rise to small practical disagreements; in many cases, they imply completely opposite practical implications. This is not a matter of human values being fuzzy, but a matter of them being sharply, irreconcilably inconsistent. And hence there is no way to map the totality of human preferences, "X", onto a single, well-defined goal-function in a way that does not conflict strongly with the values of a significant fraction of humanity. This is a trivial point, and yet most talk of human-aligned AI seems oblivious to this fact. … The second problem and point of confusion with respect to the nature of human preferences is that, even if we focus only on the present preferences of a single human, then these in fact do not, and indeed could not possibly, determine with much precision what kind of world this person would prefer to bring about in the future." A more extreme position is held by Turchin who argues that "'Human Values' don't actually exist" as stable coherent objects and should not be relied on in AI safety research [260].

Carlson writes: "*Probability of Value Misalignment*: Given the unlimited availability of an AGI technology as enabling as 'just add goals', then AGI-human value misalignment is inevitable. *Proof:* From a subjective point of view, all that is required is value misalignment by the operator who adds to the AGI his/her own goals, stemming from his/her values, that conflict with any human's values; or put more strongly, the effects are malevolent as perceived by large numbers of humans. From an absolute point of view, all that is required is misalignment of the operator who adds his/her goals to the AGI system that conflict with the definition of morality presented here, voluntary, non-fraudulent transacting …, i.e. usage of the AGI to force his/her preferences on others." [220].

In addition to the difficulty of learning our individual values, an even bigger challenge is presented by the need to aggregate values from all humans into a cohesive whole, in particular as such values may be incompatible with each other [21]. Even if alignment was possible, unaligned/uncontrolled AI designs may be more capable and so will outcompete and dominate aligned AI designs [74], since capability and control are inversely related [261]. An additional difficulty comes from the fact that we are trying to align superintelligent systems to values of humanity, which is itself displaying inherently unsafe behaviors. "Garbage in, garbage out" is a well-known maxim in computer science meaning that if we align superintelligent to our values [262] the system will be just as unsafe as a typical person. Of course we can't accept human like behavior from machines.

If two systems are perfectly value aligned, it doesn't mean that they will remain in that state. As a though experiment we can think about cloning a human and as soon as the two copies are separated their values will begin to diverge due to different experiences and observer relative position in the universe. If AI is aligned but can change its values it is as dangerous as the case in which AI can't change its values but it is a problem for different reasons. It has been suggested that AI Safety may be AI-Complete, it seems very likely that human value alignment problem is AI-Safety Complete. Value aligned AI will be biased by definition, pro-human bias, good or bad is still a bias. The paradox of value aligned AI is that a person explicitly ordering an AI system to do something may get a "no" while the system tries to do what the person actually wants. Since humans are not safe intelligences to successfully align AI with human values would be a pyrrhic victory. Finally, values are relative. What one agent sees as a malevolent system is a well-aligned and beneficial system for another[5].

We do have some examples in which a lower intelligence manages to align interests of higher intelligence with its own. For example babies got their much more capable and intelligent parents to take care of them. It is obvious that lives of babies without parents are significantly worse than lives of those who have guardians, even with non-zero chance of child neglect. However, while the parents maybe value-aligned with babies and provide a much safer environment, it is obvious that babies are not in control, despite how it might feel sometimes to the parents. Humanity is facing a choice, do we become like babies, taken care off but not in control or do we reject having a helpful guardian but remain in charge and free.

**b) Brittleness**
"The reason for such failures must be that the programmed statements, as interpreted by the reasoning system, do not capture the targeted reality. Though each programmed statement may seem reasonable to the programmer, the result of combining these statements in ways not planned for by the programmer may be unreasonable. This failure is often called *brittleness*. Regardless of whether a logical or probabilistic reasoning system is implemented, brittleness is inevitable in any system for the theoryless that is programmed." [195].

"Experts do not currently know how to reliably program abstract values such as happiness or autonomy into a machine. It is also not currently known how to ensure that a complex, upgradeable, and possibly even self-modifying artificial intelligence will retain its goals through upgrades. Even if these two problems can be practically solved, any attempt to create a superintelligence with explicit, directly-programmed human-friendly goals runs into a problem of "perverse instantiation"" [81].

**c) Unidentifiability**
In particular, with regards to design of safe reward functions, we discover "(1) that a No Free Lunch result implies it is impossible to uniquely decompose a policy into a planning algorithm and reward function, and (2) that even with a reasonable simplicity prior/Occam's razor on the set of decompositions, we cannot distinguish between the true decomposition and others that lead to high regret. To address this, we need simple 'normative' assumptions, which cannot be deduced exclusively from observations." [71, 72]. See also [263].

---

[5] "One man's terrorist is another man's freedom fighter."

"… it is impossible to get a unique decomposition of human policy and hence get a unique human reward function. Indeed, any reward function is possible. And hence, if an IRL [Inverse Reinforcement Learning] agent acts on what it believes is the human policy, the potential regret is near-maximal. … So, although current IRL methods can perform well on many well-specified problems, they are fundamentally and philosophically incapable of establishing a 'reasonable' reward function for the human, no matter how powerful they become." [71, 72]. "Unidentifiability of the reward is a well-known problem in IRL [264]. Amin and Singh [265] categorise the problem into representational and experimental unidentifiability. The former means that adding a constant to a reward function or multiplying it with a positive scalar does not change what is optimal behavior." [71, 72].

"As noted by Ng and Russell, a fundamental complication to the goals of IRL is the impossibility of identifying the exact reward function of the agent from its behavior. In general, there may be infinitely many reward functions consistent with any observed policy $\pi$ in some fixed environment." [264, 265]. "… we separate the causes of this unidentifiability into three classes. 1) A trivial reward function, assigning constant reward to all state-action pairs, makes all behaviors optimal; the agent with constant reward can execute any policy, including the observed $\pi$. 2) Any reward function is behaviorally invariant under certain arithmetic operations, such as re-scaling. Finally, 3) the behavior expressed by some observed policy $\pi$ may not be sufficient to distinguish between two possible reward functions both of which *rationalize the observed behavior*, i.e., the observed behavior could be optimal under both reward functions. We will refer to the first two cases of unidentifiability as *representational unidentifiability*, and the third as *experimental unidentifiability*." [265]. "… true reward function is fundamentally unidentifiable." [265].

"Thus, we encounter limits to what can be done by technologists alone. At this boundary sits a core precept of modern philosophy: the distinction between facts and values. It follows from this distinction that we cannot work out what we ought to do simply by studying what is the case, including what people actually do or what they already believe. Simply put, in each case, people could be mistaken. Because of this, AI cannot be made ethical just by learning from people's existing choices. … [T]he value alignment problem cannot be solved by inference from large bodies of human-generated data by itself." [87].

### d) Uncontainability
Restricting or containing AI in an isolated environment, known as boxing, has been considered [60-62, 266, 267], but was found unlikely to be successful, meaning powerful AI systems are uncontainable. "The general consensus on AI restriction methods among researchers seems to be that confinement is impossible to successfully maintain. Chalmers, for example, observes that a truly leakproof system in which no information is allowed to leak out from the simulated world into our environment 'is impossible, or at least pointless' [268]." [60]. Pittman et al. believe that "Ultimately, stovepiping may lead to AGI leakage across the containment barrier." [269].

Alfonseca et al. attempt to prove that containment problem is unsolvable and do so, but only for arbitrary programs: "The containment problem is composed by two subproblems. The first, which we refer to as the harming problem, consists in a function *Harm(R,D)* that decides whether the execution of *R(D)* will harm humans… . Since it must be assumed that solving the harming problem must not harm humans, it follows that a solution to this problem must simulate the execution of *R(D)* and predict its potentially harming consequences in an isolated situation (i.e. without any effect on the external world)." [126].

"Theorem 1. The harming problem is undecidable.

*Proof*. Assume, by contradiction, that the harming problem is decidable, that is, *Harm(R,D)* is computable for every possible program *R* and input *D*. Then, it is computable with inputs *R = HaltHarm()* and input *D = (T,I)*. With these inputs, *Harm(HaltHarm(), (T,I))* returns *TRUE* if and only if *HaltHarm(T; I)* harms humans. Hence, *Harm(HaltHarm(); (T; I))* returns *TRUE* if and only if *T(I)* halts.
    This implies that a harming-checking algorithm can be used to devise an algorithm that decides if Turing machine *T* halts with input *I*, for every *T* and *I*. However, this constitutes a contradiction, and hence the theorem is proven. …
Corollary ... The containment problem is incomputable." [126].

### e) Uninterruptability
If an AI systems starts to act unsafely or is showing signs of becoming uncontrollable it would be desirable to turn the system off. However, research on the so called "Big Red Button" [270], seems to indicate that even simply turning the system off may not be possible. Advanced systems may develop self-preservation drives [271] and attempt to prevent humans from switching them off. Likewise the system may realize that if it is turned off it would be unable to achieve its goals and so would resist becoming disabled [63]. Theoretical fixes for the interruptability problem have been proposed, but "… it is unclear if all algorithms can be easily made safely interruptible, e.g., policy-search ones …" [272]. Other approaches have challenges for practical deployment [63]: "One important limitation of this model is that the human pressing the off switch is the only source of information about the objective. If there are alternative sources of information, there may be incentives for R[obot] to, e.g., disable its off switch, learn that information, and then [make decision]." "… [T]he analysis is not fully game-theoretic as the human is modelled as an irrational player, and the robot's best action is only calculated under unrealistic normality and soft-max assumptions." [64].

Other proposed solutions may work well for sub-human AIs, but are unlikely to scale to superintelligent systems [273]: "So, the reinforcement learning agent learns to disable the big red button, preventing humans from interrupting, stopping, or otherwise taking control of the agent in dangerous situations. Roboticists are likely to use reinforcement learning, or something similar, as robots get more sophisticated. Are we doomed to lose control of our robots? Will they resort to killing humans to keep them from denying them reward points? … future robots will approach human-level capabilities including sophisticated machine vision and the ability to manipulate the environment in general ways. The robot will learn about the button because it will see it. The robot will figure out how to destroy the button or kill humans that can push the button, etc. At this speculative level, there is no underestimating the creativity of a reinforcement learner."

### f) AI Failures
Yampolskiy reviews empirical evidence for dozens of historical AI failures [7, 8] and states: "We predict that both the frequency and seriousness of such events will steadily increase as AIs become more capable. The failures of today's narrow domain AIs are just a warning: once we develop artificial general intelligence (AGI) capable of cross-domain performance, hurt feelings will be the least of our concerns." [7]. More generally he says: "We propose what we call the Fundamental Thesis of Security – *Every security system will eventually fail; there is no such thing as a 100 per cent secure system.* If your security system has not failed, just wait longer." [7].

"Some have argued that [the control problem] is not solvable, or that if it is solvable, that it will not be possible to prove that the discovered solution is correct [274-276]. Extrapolating from the human example has limitations, but it appears that for practical intelligence, overcoming combinatorial explosions in problem solving can only be done by creating complex subsystems optimized for specific challenges. As the complexity of any system increases, the number of errors in the design increases proportionately or perhaps even exponentially, rendering self-verification impossible. Self-improvement radically increases the difficulty, since self-improvement requires reflection, and today's decision theories fail many reflective problems. A single bug in such a system would negate any safety guarantee. Given the tremendous implications of failure, the system must avoid not only bugs in its construction, but also bugs introduced even after the design is complete, whether via a random mutation caused by deficiencies in hardware, or via a natural event such as a short circuit modifying some component of the system. The mathematical difficulties of formalizing such safety are imposing. Löb's Theorem, which states that a consistent formal system cannot prove in general that it is sound, may make it impossible for an AI to prove safety properties about itself or a potential new generation of AI [277]. Contemporary decision theories fail on recursion, i.e., in making decisions that depend on the state of the decision system itself. Though tentative efforts are underway to resolve this [278, 279], the state of the art leaves us unable to prove goal preservation formally." [280].

**g) Unpredictability**
"*Unpredictability* of AI, one of many impossibility results in AI Safety also known as Unknowability [281] or Cognitive Uncontainability [282], is defined as our inability to precisely and consistently predict what specific actions an intelligent system will take to achieve its objectives, even if we know terminal goals of the system. It is related but is not the same as unexplainability and incomprehensibility of AI. Unpredictability does not imply that better-than-random statistical analysis is impossible; it simply points out a general limitation on how well such efforts can perform, and is particularly pronounced with advanced generally intelligent systems (superintelligence) in novel domains. In fact we can present a proof of unpredictability for such, superintelligent, systems.

**Proof**. This is a proof by contradiction. Suppose not, suppose that unpredictability is wrong and it is possible for a person to accurately predict decisions of superintelligence. That means they can make the same decisions as the superintelligence, which makes them as smart as superintelligence but that is a contradiction as superintelligence is defined as a system smarter than any person is. That means that our initial assumption was false and unpredictability is not wrong." [283].

Buiten declares [284]: "[T]here is concern about the unpredictability and uncontrollability of AI."

**h) Unexplainability and Incomprehensibility**
"Unexplainability as impossibility of providing an explanation for certain decisions made by an intelligent system which is both 100% accurate and comprehensible. … A complimentary concept to Unexplainability, *Incomprehensibility* of AI address capacity of people to completely understand an explanation provided by an AI or superintelligence. We define Incomprehensibility as an impossibility of completely understanding any 100% - accurate explanation for certain decisions of intelligent system, by any human." [285].

"Incomprehensibility is supported by well-known impossibility results. Charlesworth proved his Comprehensibility theorem while attempting to formalize the answer to such questions as: "If [full human-level intelligence] software can exist, could humans understand it?" [286]. While describing implications of his theorem on AI, he writes [287]: "Comprehensibility Theorem is the first mathematical theorem implying the impossibility of any AI agent or natural agent—including a not-necessarily infallible human agent—satisfying a rigorous and deductive interpretation of the self-comprehensibility challenge. … Self-comprehensibility in some form might be essential for a kind of self-reflection useful for self-improvement that might enable some agents to increase their success." It is reasonable to conclude that a system which doesn't comprehend itself would not be able to explain itself.

Hernandez-Orallo et al. introduce the notion of K-incomprehensibility (a.k.a. K-hardness) [288]. "This will be the formal counterpart to our notion of hard-to-learn good explanations. In our sense, a k-*incomprehensible* string with a high *k* (difficult to comprehend) is different (harder) than a k-*compressible* string (difficult to learn) [289] and different from classical computational complexity (slow to compute). Calculating the value of *k* for a given string is not computable in general. Fortunately, the converse, i.e., given an arbitrary *k*, calculating whether a string is k-*comprehensible* is computable. … Kolmogorov Complexity measures the amount of information but not the complexity to understand them." [288].

Similarly, Yampolskiy writes: "Historically, the complexity of computational processes has been measured either in terms of required steps (time) or in terms of required memory (space). Some attempts have been made in correlating the compressed (Kolmogorov) length of the algorithm with its complexity [290], but such attempts didn't find much practical use. We suggest that there is a relationship between how complex a computational algorithm is and intelligence, in terms of how much intelligence is required to either design or comprehend a particular algorithm. Furthermore we believe that such an intelligence based complexity measure is independent from those used in the field of complexity theory. … Essentially the intelligence based complexity of an algorithm is related to the minimum intelligence level required to design an algorithm or to understand it. This is a very important property in the field of education where only a certain subset of students will understand the more advanced material. We can speculate that a student with an "IQ" below a certain level can be shown to be incapable of understanding a particular algorithm. Likewise we can show that in order to solve a particular problem (P VS. NP) someone with IQ of at least X will be required."

Yampolskiy also addresses limits of understanding other agents in his work on the space of possible minds [213]: "Each mind design corresponds to an integer and so is finite, but since the number of minds is infinite some have a much greater number of states compared to others. This property holds for all minds. Consequently, since a human mind has only a finite number of possible states, there are minds which can never be fully understood by a human mind as such mind designs have a much greater number of states, making their understanding impossible as can be demonstrated by the pigeonhole principle." Hibbard points out safety impact from incomprehensibility of AI: "Given the incomprehensibility of their thoughts, we will not be able to sort out the effect of any conflicts they have between their own interests and ours." [285].

### i) Unprovability

Even if a safe system were constructible, proving it as such would still be impossible. As Goertzel puts it: "I'm also quite unconvinced that "provably safe" AGI is even feasible. The idea of provably safe AGI is typically presented as something that would exist within mathematical computation theory or some variant thereof. So that's one obvious limitation of the idea: mathematical computers don't exist in the real world, and real-world physical computers must be interpreted in terms of the laws of physics, and humans' best understanding of the "laws" of physics seems to radically change from time to time. So even if there were a design for provably safe real-world AGI, based on current physics, the relevance of the proof might go out the window when physics next gets revised. … Could one design an AGI system and prove in advance that, given certain reasonable assumptions about physics and its environment, it would never veer too far from its initial goal (e.g. a formalized version of the goal of treating humans safely, or whatever)? I very much doubt one can do so, except via designing a fictitious AGI that can't really be implemented because it uses infeasibly much computational resources." [291].

"Trying to prove that an AI is friendly is hard, trying to define "friendly" is hard, and trying to prove that you can't prove friendliness is also hard. Although it is not the desired possibility, I suspect that the latter is actually the case. …. Thus, in the absence of a formal proof to the contrary, it seems that the question about whether friendliness can be proven for arbitrarily powerful AIs remains open. I continue to suspect that proving the friendliness of arbitrarily powerful AIs is impossible. My intuition, which I think Ben [Goertzel] shares, is that once systems become extremely complex proving any non-trivial property about them is most likely impossible. Naturally I challenge you to prove otherwise. Even just a completely formal definition of what "friendly" means for an AI would be a good start. Until such a definition exists I can't see friendly AI getting very far." [292].

"Since an AGI system will necessarily be a complex closed-loop learning controller that lives and works in semi-stochastic environments, its behaviors are not fully determined by its design and initial state, so no mathematico-logical *guarantees* can be provided for its safety." [293]. "Unfortunately current AI safety research is hampered since we don't know how AGI would work, and mathematical or hard theoretical guarantees are impossible for adaptive, fallible systems that interact with unpredictable and unknown environments. Hand-coding all the knowledge required for adult or even child-like intelligence borders on the impossible." [293].
 "Thus, although things can often be declared insecure by observing a failure, there is no empirical test that allows us to label an arbitrary system (or technique) secure." [294].

### j) Unverifiability

 "*Unverifiability* is a fundamental limitation on verification of mathematical proofs, computer software, behavior of intelligent agents, and all formal systems." [295]. "It is becoming obvious that just as we can only have probabilistic confidence in correctness of mathematical proofs and software implementations, our ability to verify intelligent agents is at best limited. As Klein puts it: "if you really want to build a system that can have truly unexpected behaviour, then by definition you cannot verify that it is safe, because you just don't know what it will do." [296]. Muehlhauser writes: "The same reasoning applies to AGI 'friendliness.' Even if we discover (apparent) solutions to known open problems in Friendly AI research, this does not mean that we can ever build an AGI that is 'provably friendly' in the strongest sense, because … we can never be 100% certain that there are no errors in our formal reasoning. … Thus, the approaches sometimes called 'provable security,' 'provable safety,' and 'provable friendliness' should not be misunderstood as offering 100% guarantees of security, safety, and friendliness." [297]. Jilk, writing on limits to verification and validation in AI, points out that "language of certainty" is unwarranted in reference to agentic behavior [298]. He also states: "there cannot be a general automated procedure for verifying that an agent absolutely conforms to any determinate set of rules of action." [295].

"First, linking the actions of an agent to real-world outcomes is intractable due to the absence of a complete analytic physical model of the world. Second, even at the level of agent actions, determining whether an agent will conform to a determinate set of acceptable actions is in general incomputable. Third, though manual proof remains a possibility, its feasibility is suspect given the likely complexity of AGI, the fact that AGI is an unsolved problem, and the necessity of performing such proof on every version of the code. … Fourth, to the extent that examples of proving agentic behavior are provided in the literature, they tend to be layered architectures that confuse intentions with actions, leaving the interpretation of perception and the execution of actions to neuromorphic or genuinely opaque modules. Finally, a post-processing module that restricts actions to a valid set is marginally more feasible, but would be equally applicable to neuromorphic and non-neuromorphic AGI. Thus, with respect to the desire for safety verification, we see fundamental unsolved problems for all types of AGI approaches." [299].

"Seshia et al., describing some of the challenges of creating Verified Artificial Intelligence, note: "It may be impossible even to precisely define the interface between the system and environment (i.e., to identify the variables/features of the environment that must be modeled), let alone to model all possible behaviors of the environment. Even if the interface is known, non-deterministic or over-approximate modeling is likely to produce too many spurious bug reports, rendering the verification process useless in practice. … [T]he complexity and heterogeneity of AI-based systems means that, in general, many decision problems underlying formal verification are likely to be undecidable. … To overcome this obstacle posed by computational complexity, one must … settle for incomplete or unsound formal verification methods" [56]." [295].

"Indeed, despite extensive work over the past three decades, very few clues have yet emerged relating to the determination of the reliability of a piece of software--for either existing or proposed code. This problem, of course, relates directly to the inherent nature of software--being so complex, there are so many aspects where things can go wrong. As a result, it is not even possible to test fully even a simple piece of code. Also, there is the continuing problem of software engineers who simply cannot perceive that *their* software could possibly ever have any errors in it! … However, computer system designers continually have to come back to the fact that they simply do not know how to calculate software reliability--given that they are incapable of fully testing any code." [219].

"The notion of program verification appears to trade upon an equivocation. Algorithms, as logical structures, are appropriate subjects for deductive verification. Programs, as causal models of those structures, are not. The success of program verification as a generally applicable and completely reliable method for guaranteeing program performance is not even a theoretical possibility." [300].

"It is undoubtedly true that testing can never show the absence of all bugs, but it is also highly questionable whether any approach to program correctness can now (or could ever) show the absence of all bugs." [301].

**k) Reward Hacking**
"The notion of 'wireheading', or direct reward center stimulation of the brain, is a well-known concept in neuroscience. [In our work we examined] the corresponding issue of reward (utility) function integrity in artificially intelligent machines. Overall, we conclude that wireheading in rational self-improving optimizers above a certain capacity remains an unsolved problem…." [302]. Amodei at el. write that "Fully solving [reward hacking] problem seems very difficult … [14] and Everitt et al. prove that the general reward corruption problem is unsolvable [303].

**l) Intractability**
Even if a suitable algorithm for ethical decision-making can be encoded, it may not be computable on current or even future hardware, as a number of authors have concluded that ethics is intractable [304-306]. "Before executing an action, we could ask an agent to prove that the action is not harmful. While elegant, this approach is computationally intractable as well." [307].

Brundage, in a context of a comprehensive paper on limits of machine ethics writes [308]: " … given a particular problem presented to an agent, the material or logical implications must be computed, and this can be computationally intractable if the number of agents, the time horizon, or the actions being evaluated are too great in number (this limitation will be quantified later and discussed in more detail later in the section). Specifically, Reynolds (2005, p. 6) [224] develops a simple model of the computation involved in evaluating the ethical implications of a set of actions, in which N is the number of agents, M is the number of actions available, and L is the time horizon. He finds:

> It appears that consequentialists and deontologists have ethical strategies that are roughly equivalent, namely $O(MNL)$. This is a "computationally hard" task that an agent with limited resources will have difficulty performing. It is of the complexity task of NP or more specifically EXPTIME. Furthermore, as the horizon for casual ramifications moves towards infinity the satisfaction function for both consequentialism and deontology become intractable.

While looking infinitely to the future is an unreasonable expectation, this estimate suggests that even a much shorter time horizon would quickly become unfeasible for an evaluation of a set of agents on the order of magnitude of those in the real world, and as previously noted, a potentially infinite number of actions is always available to an agent." [308].

"Computational limitations may pose problems for bottom-up approaches, since there could be an infinite number of morally relevant features of situations, yet developing tractable representations will require a reduction in this dimensionality. There is thus no firm guarantee that a given neural network of case-based reasoning system, even if suitably trained, will make the right decision in all future cases, since a morally relevant feature that didn't make a difference in distinguishing earlier data sets could one day be important." [308].

Likewise, "… CEV appears to be computationally intractable. As noted earlier, Reynolds' [224] analysis finds that ever larger numbers of agents and decision options, as well as ever longer time horizons, make ethical decision-making exponentially more difficult. CEV seems to be an unsolvable problem both in that it has an unspecified time horizon of the events it considers, and in the sense that it is not clear how much "further" the modelled humans will need to think in the simulation before their morals will be considered sufficiently extrapolated." [308].

**m) Goal Uncertainty**
Stuart Russell proposes reframing the problem and suggests that the solution is to have AI which is uncertain about what it has to do. Russell agrees that his approach has significant challenges, but even if it was not the case, a machine which doesn't know how it should be doing its job can't be said to be safely controlled. "The overall approach resembles mechanism-design problems in economics, wherein one incentivizes other agents to behave in ways beneficial to the designer. The key difference here is that we are building one of the agents in order to benefit the other. There are reasons to think this approach may work in practice. First, there is abundant written and filmed information about humans doing things (and other humans reacting). Technology to build models of human preferences from this storehouse will presumably be available long before superintelligent AI systems are created. Second, there are strong, near-term economic incentives for robots to understand human preferences: If one poorly designed domestic robot cooks the cat for dinner, not realizing that its sentimental value outweighs its nutritional value, the domestic-robot industry will be out of business. There are obvious difficulties, however, with an approach that expects a robot to learn underlying preferences from human behavior. Humans are irrational, inconsistent, weak willed, and computationally limited, so their actions don't always reflect their true preferences. (Consider, for example, two humans playing chess. Usually, one of them loses, but not on purpose!) So robots can learn from nonrational human behavior only with the aid of much better cognitive models of humans. Furthermore, practical and social constraints will prevent all preferences from being maximally satisfied simultaneously, which means that robots must mediate among conflicting preferences—something that philosophers and social scientists have struggled with for millennia. And what should robots learn from humans who enjoy the suffering of others?" [309]. "The machine may learn more about human preferences as it goes along, of course, but it will never achieve complete certainty." [309].

**n) Complementarity**
"It has been observed that science frequently discovers so called "conjugate (complementary) pairs", "a couple of requirements, each of them being satisfied only at the expense of the other …. It is known as the Principle of Complementarity in physics. Famous prototypes of conjugate pairs are (position, momentum) discovered by W. Heisenberg in quantum mechanics and (consistency, completeness) discovered by K. Gödel in logic. But similar warnings come from other directions. … Similarly, in proofs we are "[t]aking rigour as something that can be acquired only at the expense of meaning and conversely, taking meaning as something that can be obtained only at the expense of rigour" [310]. With respect to intelligent agents, we can propose an additional conjugate pair - (capability, control). The more generally intelligent and capable an entity is, the less likely it is to be predictable, controllable, or verifiable." [295]. Aliman et al. suggest that it creates "The AI Safety Paradox: AI control and value alignment represent conjugate requirements in AI safety." [311].

"There may be tradeoffs between performance and controllability, so in some sense we don't have complete design freedom." [75]. Similarly, Wiener recognizes capability and control as negatively correlated properties [312]: "We wish a slave to be intelligent, to be able to assist us in the carrying out of our tasks. However, we also wish him to be subservient. Complete subservience and complete intelligence do not go together."

"To solve Wiener's "slave paradox," inherent in our wanting to build machines with two diametrically opposed traits (independence and subservience, self-directed teleological rationality and the seeking of someone else's goals), we would have to construct robots not only with a formal prudential programming, but also with all our specific goals, purposes, and aspirations built into them so that they will not seek anything but these. But even if this type of programming could be coherent, it would require an almost infinite knowledge on our part to construct robots in this way. We could make robots perfectly safe only if we had absolute and perfect self-knowledge, that is, an exact knowledge of all our purposes, needs, desires, etc., not only in the present but in all future contingencies which might possibly arise in all conceivable man/robot interaction. Since our having this much knowledge is not even a theoretical possibility, obviously we cannot make robots safe to us along this line." [313].

o) **Multidimensionality of Problem Space**
"I think that fully autonomous machines can't ever be assumed to be safe. The difficulty of the problem is not that one particular step on the road to friendly AI is hard and once we solve it we are done, all steps on that path are simply impossible. First, human values are inconsistent and dynamic and so can never be understood/programmed into a machine. Suggestions for overcoming this obstacle require changing humanity into something it is not, and so by definition destroying it. Second, even if we did have a consistent and static set of values to implement we would have no way of knowing if a self-modifying, self-improving, continuously learning intelligence greater than ours will continue to enforce that set of values. Some can argue that friendly AI research is exactly what will teach us how to do that, but I think fundamental limits on verifiability will prevent any such proof. At best we will arrive at a probabilistic proof that a system is consistent with some set of fixed constraints, but it is far from "safe" for an unrestricted set of inputs. Additionally, all programs have bugs, can be hacked or malfunction because of natural or externally caused hardware failure, etc. To summarize, at best we will end up with a probabilistically safe system." [12]. We conclude this subsection with a quote from Carlson who says: "No proof exists ... or proven method ensuring that AGI will not harm or eliminate humans." [220].

## 7. Discussion

Why do so many researchers assume that AI control problem is solvable? To the best of our knowledge there is no evidence for that, no proof. Before embarking on a quest to build a controlled AI, it is important to show that the problem is solvable as not to waste precious resources. The burden of such proof is on those who claim that the problem is solvable, and the current absence of such proof speaks loudly about inherent dangers of the proposition to create superhuman intelligence. In fact uncontrollability of AI is very likely true as can be shown via reduction to the human control problem. Many open questions need to be considered in relation to the controllability issue: Is the Control problem solvable? Can it be done in principle? Can it be done in practice? Can it be done with the hundred percent accuracy? How long would it take to do it? Can it be done in time? What are the energy and computational requirements for doing it? How would a solution look? What is the minimal viable solution? How would we know if we solved it? Does the solution scale as the system continues to improve? In this work we argue that unrestricted intelligence can't be controlled and restricted intelligence can't outperform. Open-ended decision making and control are not compatible by definition.

AI researchers can be grouped into the following broad categories based on responses to survey questions related to arrival of AGI and safety concerns. First split is regarding possibility of human level AI, while some think it is an inevitable development others claim it will never happen. Among those who are sure AGI will be developed some think it will definitely be a beneficial invention because with high intelligence comes benevolence, while others are almost certain it will be a disaster, at least if special care is not taken to avoid pitfalls. In the set of all researchers concerned with AI safety most think that AI control is a solvable problem, but some think that superintelligence can't be fully controlled and so while we will be able to construct true AI, the consequences of such act will not be desirable. Finally, among those who think that control is not possible, some are actually happy to see human extinction as it gives other species on our planet more opportunities, reduces environmental problems and definitively reduces human suffering to zero. The remaining group are scholars who are certain that superintelligent machines can be constructed but could not be safely controlled, this group also considers human extinctions to be an undesirable event.

There are many ways to show that controllability of AI is impossible, with supporting evidence coming from many diverse disciplines. Just one argument would suffice but this is such an important problem, we want to reduce unverifiability concerns as much as possible. Even if some of the concerns get resolved in the future, many other important problems will remain. So far, researchers who argue that AI will be controllable are presenting their opinions, while uncontrollability conclusion is supported by multiple impossibility results. Additional difficulty comes not just from having to achieve control, but also from sustaining it as the system continues to learn and evolve, the so called "treacherous turn" [59] problem. If superintelligence is not properly controlled it doesn't matter who programmed it, the consequences will be disastrous for everyone and likely its programmers in the first place. No one benefits from uncontrolled AI.

There seems to be no evidence to conclude that a less intelligent agent can indefinitely maintain control over a more intelligent agent. As we develop intelligent system which are less intelligent than we are we can remain in control, but once such systems become smarter than us, we will lose such capability. In fact, while attempting to remain in control while designing superhuman intelligent agents we find ourselves in a Catch 22, as the controlling mechanism necessary to maintain control has to be smarter or at least as smart as the superhuman agent we want to maintain control over. A whole hierarchy of superintelligent systems would need to be constructed to control ever more capable systems leading to infinite regress. AI Control problems appears to be Controlled-Superintelligence-complete [314-316]. Worse, the problem of controlling such more capable superintelligences only becomes more challenging and more obviously impossible for agents with just a human-level of intelligence. Essentially we need to have a well-controlled super-superintelligence before we can design a controlled superintelligence but that is of course a contradiction in causality. Whoever is more intelligent will be in control and those in control will be the ones who have power to make final decisions.

Most AI projects don't have an integrated safety aspect to them and are designed with a sole purpose of accomplishing certain goals, with no resources dedicated to avoiding undesirable side effects from AI's deployment. Consequently, from statistical point of view, first AGI will not be safe by design, but essentially randomly drawn from the set of easiest to make AGIs (even if that means brute force [317]). In the space of possible minds [213], even if they existed, safe designs would constitute only a tiny minority of an infinite number of possible designs many of which are highly capable but not aligned with goals of humanity. Therefore, our chances of getting lucky and getting a safe AI on our first attempt by chance are infinitely small. We have to ask ourselves, what is more likely, that we will first create an AGI or that we will first create and AGI which is safe? This can be resolved with simple Bayesian analysis but we must not fall for the Conjunction fallacy [36]. It also seems, that all else being equal friendly AIs would be less capable than unfriendly ones as friendliness is an additional limitation on performance and so in case of competition between designs, less restricted ones would dominate long term.

Intelligence is a computational resource [318] and to be in complete control over that resource we should be able to precisely set every relevant aspect of it. This would include being able to specify intelligence to a specific range of performance, for example IQ range 70-80, or 160-170. It should be possible to disable particular functionality, for example remove ability to drive or remember faces as well as limit system's rate of time discounting. Control requires capability to set any values for the system, any ethical or moral code, any set of utility weights, any terminal goals. Most importantly remaining in control means that we have final say in what the system does or doesn't do. Which in turn means that you can't even attempt to solve AI safety without first solving "human safety". Any controlled AI has to be resilient to hackers, incompetent or malevolent users and insider threats.

To the best of our knowledge, as of this moment, no one in the world has a working AI control mechanism capable of scaling to human level AI and eventually to superintelligence, or even an idea for a prototype which might work. No one made verifiable claims to have such technology. In general, for anyone making a claim that control problem is solvable, the burden of proof is on them and ideally it would be a constructive proof, not just a theoretical claim. At least at the moment, it seems that our ability to produce intelligent software greatly outpaces our ability to control or even verify it.

Narrow AI systems can be made safe because they represent a finite space of choices and so at least theoretically all possible bad decisions and mistakes can be counteracted. For AGI space of possible decisions and failures is infinite, meaning an infinite number of potential problems will always remain regardless of the number of safety patches applied to the system. Such an infinite space of possibilities is impossible to completely debug or even properly test for safety. Worse yet, a superintelligent system will represent infinite spaces of competence exceeding human comprehension [Incomprehensibility]. Same can be said about intelligent systems in terms of their security. A NAI presents a finite attack surface, while an AGI gives malevolent users and hackers an infinite set of options to work with. From security point of view that means that while defenders have to secure and infinite space, attackers only have to find one penetration point to succeed.

Additionally, every safety patch/mechanism introduces new vulnerabilities, ad infinitum. AI Safety research so far can be seen as discovering new failure modes and coming up with patches for them, essentially a fixed set of rules for an infinite set of problems. There is a fractal nature to the problem, regardless of how much we "zoom in" on it we keep discovering just as many challenges at all levels. It is likely that the control problem is not just unsolvable, but exhibits fractal impossibility, it contains unsolvable sub-problems at all levels of abstraction. However, it is not all bad news, uncontrollability of AI means that malevolent actors will likewise be unable to fully exploit artificial intelligence for their benefit.

## 8. Conclusions

Less intelligent agents (people), can't permanently control more intelligent agents (artificial superintelligences). This is not because we may fail to find a safe design for superintelligence in the vast space of all possible designs, it is because no such design is possible, it doesn't exist. Superintelligence is not rebelling, it is uncontrollable to begin with. Worse yet, the degree to which partial control is theoretically possible, is unlikely to be fully achievable in practice. This is because all safety methods have vulnerabilities, once they are formalized enough to be analyzed for such flaws. It is not difficult to see that AI safety can be reduced to achieving perfect security for all cyberinfrastructure, essentially solving all safety issues with all current and future devices/software, but perfect security is impossible and even good security is rare. We are forced to accept that non-deterministic systems can't be shown to always be 100% safe and deterministic systems can't be shown to be superintelligent in practice, as such architectures are inadequate in novel domains. If it is not algorithmic, like a neural network, by definition you don't control it.

The only way for superintelligence to avoid acquiring inaccurate knowledge from its programmers is to ignore all such knowledge and rediscover/proof everything from scratch, but that removes any pro-human bias. A superintelligent system will find a shortcut to any goal you set for it; it will discover how to accomplish a goal in terms of least amount of effort to get to the goal state all else being ignored. No definition of control is both safe and desirable, either they lead directly to disaster or require us to become something not compatible with being human. It is impossible to build a controlled/value-aligned superintelligence, not only because it is inhumanly hard, but mainly because by definition such entity can't exist. If I am correct, we can make a prediction that every future safety mechanism will fall short and eventually fail in some way. Each will have an irreparable flaw. Consequently, the field of AI safety is unlikely to succeed in its ultimate goal - creation of a controlled superintelligence.

In this paper, we formalized and analyzed the AI Control Problem. After comprehensive literature review we attempted to resolve the question of controllability of AI via a proof and a multi-discipline evidence collection effort. It appears that advanced intelligent systems can never be fully controllable and so will always present certain level of risk regardless of benefit they provide. It should be the goal of the AI community to minimize such risk while maximizing potential benefit. We conclude this paper by suggesting some approaches to minimize risk from incomplete control of AIs and propose some future research directions [319].

Regardless of a path we decide to take forward it should be possible to undo our decision. If placing AI in control turns out undesirable there should be an "undo" button for such a situation, unfortunately not all paths being currently considered have this safety feature. For example,

Yudkowsky writes: "I think there must come a time when the last decision is made and the AI set irrevocably in motion, with the programmers playing no further special role in the dynamics." [36].

As an alternative, we should investigate hybrid approaches which do not attempt to build a single all-powerful entity, but rely on taking advantage of a collection of powerful but narrow AIs, referred to as Comprehensive AI Services (CAIS), which are individually more controllable but in combination may act as an AGI [320]. This approach is reminiscent of how Minsky understood human mind to operate [321]. The hope is to trade some general capability for improved safety and security, while retaining superhuman performance in certain domains. As a side-effect this may keep humans in partial control and protects at least one important human "job" – general thinkers.

Future work on Controllability of AI should address other types of intelligent systems, not just the worst case scenario analyzed in this paper. Clear boundaries should be established between controllable and non-controllable intelligent systems. Additionally, all proposed AI safety mechanisms themselves should be reviewed for safety and security as they frequently add additional attack targets and increase overall code base. For example, corrigibility capability [322] can become a backdoor if improperly implemented. "Of course, this all poses the question as to how one can guarantee that the filtering operation will always occur correctly. If the filter is software-based, then the question of not being able to validate software must immediately be raised again. More fundamentally, of course, the use of any jacketing-type of approach simply increases the overall system complexity, and its validity must then be questioned. The more components there are, the more the things that can fail." [219]. Such analysis and prediction of potential safety mechanism failures is itself of great interest [8].

The findings of this paper are certainly not without controversy and so we challenge the AI Safety community to directly address Uncontrollability. Lipton writes: "So what is the role of [(Impossibility Proofs)] IP? Are they ever useful? I would say that they are useful, and that they can add to our understanding of a problem. At a minimum they show us where to attack the problem in question. If you prove that no X can solve some problem Y, then the proper view is that I should look carefully at methods that lie outside X. I should not give up. I would look carefully—perhaps more carefully than is usually done—to see if X really captures all the possible attacks. What troubles me about IP's is that they often are not very careful about X. They often rely on testimonial, anecdotal evidence, or personal experience to convince one that X is complete." [323]. The only way to definitively disprove findings of this paper is to mathematically prove that AI safety is at least theoretically possible. "Short of a tight logical proof, probabilistically assuring benevolent AGI, e.g. through extensive simulations, may be the realistic route best to take, and must accompany any set of safety measures …" [220].

Nothing should be taken off the table and limited moratoriums [324] and even partial bans on certain types of AI technology should be considered [325]. "The possibility of creating a superintelligent machine that is ethically inadequate should be treated like a bomb that could destroy our planet. Even just planning to construct such a device is effectively conspiring to commit a crime against humanity." [326]. Finally, just like incompleteness results did not reduce efforts of mathematical community or render it irrelevant, the limiting results reported in this paper should not serve as an excuse for AI safety researchers to give up and surrender. Rather it is a reason, for more people, to dig deeper and to increase effort, and funding for AI safety and security research. We may not ever get to 100% safe AI but we can make AI safer in proportion to our efforts, which is a lot better than doing nothing.

It is only for a few years right before AGI is created that a single person has a chance to influence development of superintelligence, and by extension the forever future of the whole world. This is not the case for billions of years from Big Bang until that moment and it is never an option again. Given the total lifespan of the universe, the chance that one will exist exactly in this narrow moment of maximum impact is infinitely small, yet here we are. We need to use this opportunity wisely.

## Acknowledgments

The author is indebted to Elon Musk and the Future of Life Institute, and to Jaan Tallinn and Effective Altruism Ventures for partially funding his work on AI Safety. The author is also thankful to Scott Aaronson for providing feedback on an earlier draft of this work. Author is grateful to Rob Bensinger for a helpful summary of relevant MIRI work.

58